\documentclass[12pt]{iopart}
\usepackage{iopams}

\newcommand{\uD}[1]{\hspace{-0em}\mathcal{D}{#1}\;}
\newcommand{\uDm}[1]{\hspace{-0em}\mathcal{D}{#1}}
\newcommand{\ud}[1]{\hspace{-0em}\mathrm{d}{#1}\;}
\newcommand{\udm}[1]{\hspace{-0em}\mathrm{d}{#1}}
\newcommand{\udd}[2]{\hspace{-0em}\mathrm{d}{#1}\,\mathrm{d}{#2}\;}

\newcommand{\Ud}[1]{\hspace{-0.5ex}\mathrm{d}{#1}\;}
\newcommand{\Udm}[1]{\hspace{-0.5ex}\mathrm{d}{#1}}
\newcommand{\Udd}[2]{\hspace{-0.5ex}\hspace{-0em}\mathrm{d}{#1}\,\mathrm{d}{#2}\;}
\newcommand{\iint}{\int\hspace{-0.45em}\int}

\begin{document}

\title{Symmetries of generating functionals of Langevin
processes with colored multiplicative noise}
\author{Camille Aron$^*$, Giulio Biroli$^{\dagger}$ and Leticia F. Cugliandolo$^*$}
\address{$^*$Laboratoire de Physique Th\'eorique et Hautes \'Energies, \\
Universit\'e Pierre et Marie Curie - Paris VI, \\
4 Place Jussieu, Tour 24, 5\`eme \'etage, 75252 Paris Cedex 05, France\\
$^{\dagger}$Institut de Physique Th\'eorique, CEA Saclay, 91191 Gif-sur-Yvette, France}

\begin{abstract}
  We present a comprehensive study of the symmetries of the generating
  functionals of generic Langevin processes with multiplicative
  colored noise.  We treat both Martin-Siggia-Rose-Janssen-deDominicis
  and supersymmetric formalisms.  We summarize the relations between
  observables that they imply including fluctuation relations,
  fluctuation-dissipation theorems, and Schwinger-Dyson
  equations. Newtonian dynamics and their invariances follow in the
  vanishing friction limit.
\end{abstract}

\maketitle
\parskip=2.5pt
\tableofcontents

\newpage

\parskip = 3.5pt

\section{Introduction}
 
The stochastic evolution of a classical system coupled to a quite
generic environment can be described with the Langevin
formalism~\cite{Langevin,book-on-Langevin} and its generating
functional, the Martin-Siggia-Rose-Janssen-deDominicis (MSRJD)
path-integral~\cite{Martin-Siggia-Rose,Jenssen,deDominicis}. In many
cases of practical interest the effect of the environment is captured
by an additive white noise and its memory-less friction, Brownian
motion being the paradigmatic example~\cite{Langevin}. Nevertheless,
there are many other interesting instances in which the noise is
multiplicative and colored, and the friction effect is consistently
described by a memory kernel coupled to a non-linear function of the
state variable. Such Langevin equations appear in many different
branches of physics (as well as chemistry and other sciences).   In magnetism, the motion of the
classical magnetic moments of small particles is phenomenologically
described by the Landau-Lifshitz-Gilbert equation in which the
fluctuations of the magnetic field are coupled multiplicatively to the
magnetic moment~\cite{Gilbert}. Many other examples pertain to soft
condensed matter; two of these are confined diffusion, in which the
diffusion coefficient of the particle depends on the position {\it
  via} hydrodynamic interactions~\cite{Lubensky}, and the stochastic
partial differential equation that rules the time-evolution of the
density of an ensemble of $N$ Brownian particles~\cite{Dean}.  In a
cosmological framework, these equations arise as effective equations of motion for the
out of (although close to) equilibrium evolution of self-interacting
quantum fields in which the short-wave length modes serve as thermal
baths for longer wave-length modes with slower
dynamics~\cite{cosmological}. Such type of fluctuations may yield {\it a
  priori} unexpected results such as noise induced phase transitions
in systems in which the associated deterministic potential does not
exhibit any symmetry breaking~\cite{noise-induced}.

In order to better understand these processes it is useful to
distinguish cases in which sources of fluctuations and dissipation can
be different.  On the one hand, the noise and friction term can have
an `internal' origin, such as in diffusion problems. On the other hand,
the stochastic fluctuations can be due to an `external'
source~\cite{vanKampen}.  In the former cases one usually assumes that
the variables generating the noise and friction are in equilibrium and
the terms in the Langevin equation associated to them are linked by a
fluctuation-dissipation theorem. In the absence of non-conservative
external forces the Boltzmann measure of the system of interest is a
steady state of its dynamics. In the latter cases noise and dissipation are
not forced to satisfy any equilibrium condition and this translates
into the possibility of having any kind of noise and friction
terms.  For concreteness we shall focus on the first type of problems
and only mention a few results concerning the latter.

In treatments of the examples mentioned in the first paragraph,
the delicate double limit of vanishing fast variables relaxation time
and noise correlation time is often taken. These lead to a first order
stochastic differential equation with multiplicative white noise.  Its
interpretation in the It\^o, Stratonovich or other sense requires a very careful
analysis of the order of limits, see e.g.~\cite{Kupferman} and
references therein.  In the main part of this paper we shall keep both
time-scales finite and thus avoid the subtleties encountered in the
double vanishing limit.

In this manuscript, we identify a number of symmetries of the MSRJD generating
functional of inertial Langevin processes with multiplicative colored
noise. One symmetry is only valid in equilibrium. The corresponding 
Ward-Takahashi identities between the correlation functions of the field theory lead 
to various equilibrium relations such as stationarity, 
fluctuation-dissipation theorems~\cite{FDT,Kubo-FDT} or Onsager relations.
Away from equilibrium, the symmetry is broken giving rise to various 
out of equilibrium fluctuation relations~\cite{Evans}-\cite{FT-reviews}.
Another symmetry holds for generic out of equilibrium set-ups
and implies dynamic equations coupling correlations and linear responses.
It allows  to express, in particular, the linear response in terms of
correlations without applying a perturbing field~\cite{Cukupa}-\cite{Chetrite}.

We are aware of the fact that some of the results in this manuscript
-- especially, in the limit of additive noise -- were already known
and we do our best to attribute them to the authors of the original
papers or review articles.  Still, the presentation that we gradually
develop in this article allows one to go beyond the simple cases and
treat the multiplicative non-Markov processes with the same level
of difficulty. Moreover, we discuss in greater detail than previously
done the transformation of the measure and several Jacobians, and the
domain of integration of the fields in the path-integral. The
importance of dealing with colored noise, and to treat the
transformation of the fields in the complex plane, is enhanced by our
purpose to extend this analysis to quantum dissipative problems.
These results will be presented in a separate
publication~\cite{Arbicu}.  

The organization of the paper is the following. In Sect.~\ref{sec:langevin} we give a short review of
Langevin equations with additive and multiplicative noise. 
Section~\ref{sec:TheGF} presents the MSRJD functional representation of Langevin 
equations. In Sect.~\ref{sec:equilibrium} we deal with the equilibrium symmetries 
while in Sect.~\ref{sec:outofequilibrium} we treat the out of equilibrium ones. In both Sections we 
discuss supersymmetric formulations. We conclude in Sect.~\ref{sec:conclusions}.

\section{The Langevin equation}
\label{sec:langevin}

We consider a $0$-dimensional field $\psi$ (\textit{e.g.} a particle
at position $\psi$) with mass $m$ driven by a force $F$ and in contact
with a thermal bath in equilibrium at inverse temperature $\beta$.
The initial time, $t_0$, is the instant at which the particle is set
in contact with the bath and the stochastic dynamics `starts'. We call
it $t_0=-T$ and without loss of generality we work within a symmetric
time-interval $t \in [-T,T]$.  The extension to higher dimensional
cases is straightforward. Our conventions are given in \ref{sec:conventions}.

\subsection{Additive noise}

The Langevin equation with additive noise
is given by
\begin{eqnarray} \label{eq:Langevin}
\mbox{\sc{Eq}}([\psi],t) &\equiv& 
m \ddot \psi(t) - F([\psi],t) + \int_{-T}^{T} 
\Ud{u} \gamma(t,u) \, \dot \psi(u) 
=
\xi(t) \;,
\end{eqnarray}
with $\dot \psi(t)=\mathrm{d}\psi(t)/\mathrm{d}t$ and $\ddot \psi(t)={\rm
  d}^2\psi(t)/\mathrm{d}t^2$. The force can be decomposed into
conservative and non-conservative parts: $F([\psi],t) =
-V'(\psi(t),\lambda(t)) + \mathrm{f}([\psi],t)$. $V$ is a local potential the
time-dependence of which is controlled externally through a protocol
$\lambda(t)$. $V'$ denotes the partial derivative of $V$ with respect
to $\psi$. $\mathrm{f}([\psi],t)$ collects all the non-conservative forces that
are externally applied.
$\mathrm{f}([\psi],t)$ is assumed to be causal in the sense that it does not depend on the future states of the system, 
$\psi(t')$ with $t' > t$. Furthermore, we suppose that $\mathrm{f}([\psi],t)$ 
does not involve second -- nor higher -- order time-derivatives of the field $\psi(t)$.
The last term in the left-hand-side ({\sc
  lhs}) and the right-hand-side ({\sc rhs}) of the equation model
the interaction with the bath. 
These two heuristic terms can be derived 
using a model~\cite{Zwanzig,Weiss} 
in which the bath consists in a set of non-interacting harmonic 
oscillators of coordinates $q_i$ that are bilinearly coupled  to the 
state variable of the system of interest $\psi$.
The function $\gamma$ is the retarded
friction [$\gamma(t,t') = 0$ for $t'>t$] and the noise $\xi$ is a
random force taken to be a Gaussian process. This assumption is quite
reasonable, for instance, for a Brownian particle 
much bigger than the surrounding particles of the bath,
its motion being the result of a large number of successive collisions, which is a
condition for the central limit theorem to apply. Since we assume the
environment to be in equilibrium, $\gamma(t,t')$ is a function of
$t-t'$ and the bath obeys the fluctuation-dissipation theorem of the
`second kind'~\cite{Kubo}:
\begin{eqnarray}
 \langle \xi(t) \rangle_\xi = 0 \; , \qquad
 \langle \xi(t) \xi(t') \rangle_\xi =  \beta^{-1} \ \Gamma(t-t') 
\;,
\label{eq:noise}
\end{eqnarray}
where $\langle \, ... \, \rangle_\xi$ denotes the average over the noise
history. We introduced the symmetric kernel $\Gamma(t-t') \equiv
\gamma(t-t') + \gamma(t'-t) = \Gamma(t'-t)$.  The 
white noise limit, in which the bath has no memory, is achieved by setting
$\gamma(t-t') = \gamma_0 \delta(t-t')$ with $\gamma_0 >
0$ the friction coefficient. 
The Langevin equation then takes the more familiar form
\begin{equation} \label{eq:Langevin_white}
\mbox{\sc{Eq}}([\psi],t) \equiv 
m \ddot \psi(t) -  F([\psi],t) + \gamma_0 
\ \dot \psi(t) 
=
\xi(t)
\; . 
\end{equation}
Newtonian dynamics, for which the system is not in contact with a thermal bath, are recovered by simply taking $\gamma(t)=\Gamma(t)=0$ at all $t$.
Out of equilibrium environments can be taken into account by relaxing the condition between the noise statistics and the friction kernel $\Gamma(t-t')=\gamma(t-t')+\gamma(t'-t)$.

\subsection{Multiplicative noise}

We generalize our discussion to the multiplicative noise case in which
the Gaussian noise $\xi$ is coupled to a state-dependent function
$M'(\psi)$.  The Langevin equation reads
\begin{eqnarray}
 \label{eq:LangevinMultiplicatif}
 \mbox{\sc{Eq}}([\psi],t) &\equiv& 
m \ddot \psi(t) - F([\psi],t) + M'(\psi(t)) \int_{-T}^T \Ud{u}
 \gamma(t-u)M'(\psi(u))  \dot \psi(u) \nonumber\\
& & \qquad\qquad =
M'(\psi(t))  \, \xi(t) \;.
\label{eq:Langevin-mult}
\end{eqnarray}
This equation can also be shown by using the oscillator model for the
bath and a nonlinear coupling of the form $M(\psi)\sum_i c_i q_i$
where $c_i$ are coefficients that depend on the details of the
coupling and $M(\psi)$ is a smooth function of the state variable with
$M(0)=0$.  By a suitable renormalization of $\gamma$, one can always
achieve $M'(0) = 1$.  For reasons that will soon become clear, we
need to assume that $M'(\psi)\neq0$ $\forall\,\psi$.  These
assumptions can be realized with functions of the type $M(\psi) = \psi
+ L(\psi)$ where $L$ is a smooth and increasing function satisfying
$L(0) = L'(0)=0$.  The complicated structure of the friction term
takes its rationale from the fluctuation-dissipation theorem of the
second kind that expresses the equilibrium condition of the bath.
This equation models situations in which the friction between the
system and its bath is state-dependent. $\xi$ has the same statistics
as in the additive case, see eq.~(\ref{eq:noise}).  The Langevin
equation for the additive noise problem is recovered by taking
$M(\psi)=\psi$.

\subsection{Initial conditions}

The Langevin equation is a second order differential equation that needs two initial values, say
the field and its derivative at time $-T$. We shall use initial
conditions drawn from an initial probability distribution $P_{\rm
  i}\left(\psi(-T),\dot\psi(-T)\right)$ and average over them. The
initial conditions are not correlated with the thermal noise $\xi$. In
the particular case in which the system is prepared in an equilibrium
state, $P_\mathrm{i}$ is given by the Boltzmann measure.

\subsection{Markov limit}\label{sec:MarkovLim}

Langevin equations are often given in the Markov limit in which they appear to be
first order stochastic differential equations.
Second and higher order time-derivatives as well as non-local terms such as 
memory kernels are not allowed. In other words, the
effect of inertia is neglected (Smoluchowski limit) and the bath is
taken to be white. This is justified in situations in which the two
associated time scales are sufficiently small compared to all other
time scales involved.  Concretely, the resulting equation is derived
by using an adiabatic elimination procedure that consists in
integrating over the fast variables of the system (the velocities) and
of the bath.  However, this double limiting procedure requires a
careful analysis.

The physics of the resulting equation may depend on how the
relaxation time associated to inertia compares with the correlation
time of the noise before sending the two of them to zero.  
In cases in which the latter is much larger than the former, the limiting
stochastic equation should be interpreted in the sense of
Stratonovich~\cite{Strato}. The {\sc rhs} of eq.~(\ref{eq:LangevinMultiplicatif})
is given a meaning by stating that $\psi$ in $M'(\psi(t))$ is evaluated at half
the sum of its values before and after the kick.
Conversely, when the inertia relaxation time is much larger than the
noise correlation time, the limiting equation should be interpreted in
the It\^o sense~\cite{Ito}. In this scenario, the rule is that 
$M'(\psi(t))$ is evaluated just before the kick $\xi(t)$.
When the noise is additive the two conventions are equivalent
(see \ref{sec:MarkovianAction}) for all practical purposes. However,
they are not  for processes with multiplicative noise~\cite{vanKampen}.
For these it is possible to rewrite the It\^o stochastic equation in terms of a Stratonovich
stochastic equation by adding an adequate drift term to the
deterministic force -- and be allowed to 
use the rules of conventional calculus. 
The Fokker-Planck equation associated to the  Markov process 
does not depend on the scenario and the Boltzmann distribution is a steady state
independently of the convention used.
However, the action of the generating functional acquires extra
terms depending on the discretization prescription~\cite{Lubensky,Tirapegui}.

In this article, we decide not to cope with the Markov limit and,
unless otherwise stated, we keep the inertia of the system in
our equations ($m\neq0$) and we use a colored noise with a finite
relaxation time.

\section{The generating functional}
\label{sec:TheGF}

The generating functionals associated to the equations of motion
(\ref{eq:Langevin}) and (\ref{eq:Langevin-mult}) are given by the
Martin-Siggia-Rose-Janssen-deDominicis (MSRJD) path-integral.  In this
Section we recall its construction for additive noise~\cite{Jenssen}
and we extend it to multiplicative noise by using a continuous time
formalism.  In~\ref{app:MSRJD} and \ref{app:MSRJDmult} we develop a careful construction 
in the discretized formulation.

\subsection{Action in the additive noise case}

The Langevin equation (\ref{eq:Langevin}) is a second order
differential equation with source $\xi$. The knowledge of the history
of the field $\xi$ and the initial conditions $\psi(-T)$ and $\dot
\psi(-T)$ is sufficient to construct $\psi(t)$.  Therefore, the
probability $P[\psi]$ of a given $\psi$ history between $-T$ and $T$
is linked to the probability of the noise history $P_\mathrm{n}[\xi]$
through
\begin{eqnarray}
 P[\psi] \uDm{[\psi]} &=& P_\mathrm{n}[\xi] \ \uD{[\xi]}  P_{\rm
i}\left(\psi(-T),\dot\psi(-T)\right) \ud{\psi(-T)} \ud{\dot \psi(-T)}
\end{eqnarray}
implying
\begin{eqnarray}
 P[\psi]  
 &=& P_\mathrm{n}[\mbox{\small\sc{Eq}}[\psi]] \ \left|{\cal J}[\psi]\right| \ P_\mathrm{i}\left(\psi(-T),\dot\psi(-T)\right)\;, \label{eq:PnXI}
 \end{eqnarray}
where $\mathcal{J}[\psi]$ is the Jacobian which reads, up to some constant factor,
\begin{eqnarray}
 {\cal J}[\psi] \equiv  \mbox{det}_{uv} 
\left[ \frac{\delta\xi(u)}{\delta \psi(v)} \right]   = \mbox{det}_{uv} \left[
\frac{\delta\mbox{\small\sc{Eq}}([\psi],u)}{\delta \psi(v)} \right] \equiv {\cal J}_0[\psi]
\;.\label{eq:Jac_addi}
\end{eqnarray}
$\mbox{det} \left[ ... \right]$ stands for the functional determinant. We introduced the notation ${\cal J}_0[\psi]$ for future convenience and we shall discuss it in Sect.~\ref{sec:Jaco}.
After a Hubbard-Stratonovich transformation that introduces the auxiliary
real field $\hat\psi$, the Gaussian probability for a given noise
history to occur reads
\begin{eqnarray}
 P_\mathrm{n}[\xi]  &=& {\cal N}^{-1} 
 \int \uD{[\hat\psi]} \rme^{-\int \Ud{u} \rmi\hat\psi(u) \xi(u)  +\frac{1}{2} \iint \Udd{u}{v}
\rmi\hat\psi(u) {\beta^{-1}}\Gamma(u-v) \rmi\hat\psi(v) }\;, 
\end{eqnarray}
with the boundary conditions $\hat\psi(-T)=\hat\psi(T)=0$ and
where all the integrals over time run from $-T$ to $T$. In the
following, unless otherwise stated, we shall simply denote them
$\int$. ${\cal N}$ is a infinite constant prefactor that we absorb
in a re-definition of the measure $\uDm{[\hat\psi]}$.  Back in
eq.~({\ref{eq:PnXI}}) one has
\begin{eqnarray}
 P[\psi] \uDm{[\psi]} = \uDm{[\psi]} \int \uD{[\hat \psi]} 
\rme^{S[\psi,\hat\psi]}\;, 
\end{eqnarray}
with the MSRJD action functional
\begin{eqnarray}\label{eq:Action_Fancy}
 S[\psi, \hat\psi] &\equiv& 
\ln P_\mathrm{i}\left(\psi(-T),\dot\psi(-T)\right) 
-\int \Ud{u} \rmi\hat\psi(u) \mbox{\small\sc{Eq}}([\psi],u) \nonumber \\
 & & +\frac{1}{2} \iint \Udd{u}{v}
\rmi\hat\psi(u)\ {\beta^{-1}}\Gamma(u-v) \ \rmi\hat\psi(v) + \ln |{\cal J}_0[\psi]| \;.
\end{eqnarray}
The latter is the sum of a deterministic, a dissipative and a Jacobian term,
\begin{eqnarray}
 S[\psi, \hat\psi] &\equiv& 
S^\mathrm{det}[\psi, \hat\psi] + S^\mathrm{diss}[\psi, \hat\psi] + \ln |{\cal J}_0[\psi]| \;, \nonumber
\end{eqnarray}
with
\begin{eqnarray}
S^\mathrm{det}[\psi, \hat\psi] 
 &\equiv& 
\ln P_\mathrm{i}\left(\psi(-T),\dot\psi(-T)\right) \nonumber
\\
 & & - \int \Ud{u} \rmi\hat \psi(u) \left[ m\ddot \psi(u) 
- F([\psi],u)
\right], 
\label{eq:Sdet}
\\
S^\mathrm{diss}[\psi, \hat\psi] 
&\equiv&
\int \Ud{u} \rmi\hat \psi(u)  \int \Ud{v} \gamma(u-v) 
\left[\beta^{-1} \rmi\hat \psi(v) - \dot \psi(v)\right]\;.
\label{eq:Sdiss}
\end{eqnarray}
$S^\mathrm{det}$ takes into account inertia and the
forces exerted on the field, as well as the measure of the initial
condition.  $S^\mathrm{diss}$ has its origin in the coupling to
the dissipative bath. In the  
white noise limit, $\gamma(t-t') =
\gamma_0 \delta(t-t')$, the dissipative action naively simplifies
to $S^\mathrm{diss}[\psi, \hat\psi] = \gamma_0 \int \ud{u}
\rmi\hat \psi(u) \left[\beta^{-1} \rmi\hat \psi(u) - \dot
  \psi(u)\right]$ (see~Sect.~\ref{sec:MarkovLim} for additional details on this 
limit).

Integrating away the auxiliary field $\hat\psi$
yields the Onsager-Machlup action functional~\cite{OnsagerMachlup}. 
However, we prefer to work with the action written in terms of $\psi$ and $\rmi\hat\psi$ 
as this is the form that arises as the classical limit of the Schwinger-Keldysh 
action used to  treat interacting out of equilibrium quantum systems~\cite{Weiss,Kamenev},
that we shall analyze along the same lines in~\cite{Arbicu}.

\subsection{Action in the multiplicative noise case}

To shorten expressions, we adopt a notation in which the arguments
of the fields and functions appear as subindices, $\psi_u\equiv\psi(u)$,
$\gamma_{u-v}\equiv\gamma(u-v)$, and so on and so forth, and the
integrals over time expressed as $\int_u\equiv\int_{-T}^T \ud{u}$.

In the case of the Langevin equation~(\ref{eq:LangevinMultiplicatif})
with multiplicative noise, the relation (\ref{eq:PnXI}) is modified
and reads
\begin{eqnarray}
 P[\psi]  
 = P_\mathrm{n}\left[\frac{\mbox{\small\sc{Eq}}[\psi]}{M'(\psi)}\right] \ 
\left|{\cal J}[\psi]\right| \ P_\mathrm{i}(\psi_{-T},\dot\psi_{-T})\;,
 \end{eqnarray}
with the Jacobian 
\begin{eqnarray}
 {\cal J}[\psi] &=& \mbox{det}_{uv} \left[ \frac{\delta\mbox{\small\sc{Eq}}_u[\psi]/M'(\psi_u)}{\delta \psi_v} \right] \label{eq:new_J}
= \mbox{det}_{uv}\left[ \frac{\delta_{u-v}}{M'(\psi_u)} \right] {\cal J}_0[\psi]
\end{eqnarray}
and the generalization of 
the definition of ${\cal J}_0$ in eq.~(\ref{eq:Jac_addi}) to the multiplicative case:
\begin{eqnarray}
 {\cal J}_0[\psi] \equiv \mbox{det}_{uv} \left[ \frac{\delta\mbox{\small\sc{Eq}}_u[\psi]}{\delta \psi_v}  - \frac{M''(\psi_u)}{M'(\psi_u)} \, \mbox{\small\sc{Eq}}_u[\psi]  \, \delta_{u-v}\right] \;. \label{eq:J0_multi}
\end{eqnarray}
The construction of the MSRJD action
follows the same steps as in the additive noise case, complemented by
a further transformation of the field
$\rmi\hat\psi\mapsto\rmi\hat\psi\,M'(\psi)$, the Jacobian of which
cancels the first determinant factor in the {\sc rhs} of
eq.~(\ref{eq:new_J}). Therefore, the MSRJD action reads
\begin{eqnarray}
 S[\psi, \hat\psi] &\equiv& 
\ln P_\mathrm{i}(\psi_{-T},\dot\psi_{-T}) 
-\int_u \rmi\hat\psi_u \mbox{\small\sc{Eq}}_u[\psi] \nonumber \\
 & &\;\;\; +\frac{1}{2} \int_u \int_v
\rmi\hat\psi_u M'(\psi_u) \ {\beta^{-1}}\Gamma_{u-v} \ M'(\psi_v)\rmi\hat\psi_v
 + \ln |{\cal J}_0[\psi]|
 \;, \label{eq:MSRJDMultip}
\end{eqnarray}
with ${\cal J}_0$ defined in eq.~(\ref{eq:J0_multi}) . The deterministic part of
the action is unchanged compared to the additive noise case and the
dissipative part is now
\begin{eqnarray}
S^\mathrm{diss}[\psi, \hat\psi] 
&\equiv&
\int_u\rmi\hat \psi_u  \int_v M'(\psi_u) \, \gamma_{u-v} \, M'(\psi_v)
\left[ \beta^{-1} \rmi\hat \psi_v - \dot \psi_v \right] \label{eq:Sdiss_multi}
\;.
\end{eqnarray}

\subsection{The Jacobian} \label{sec:Jaco}

In~\ref{app:MSRJDmult} we prove that  the Jacobian ${\cal J}_0$ is a
field-independent positive constant for Langevin equations with 
inertia and multiplicative colored noise.
One can therefore safely include the Jacobian contribution 
in the normalization. However, we decide to keep track of this term 
and  represent it as a Gaussian integral over 
Grassmann conjugate fields $c$ and $c^*$,
\begin{eqnarray}
 {\cal J}_0[\psi]  =
\int \uD{[c,c^*]} \rme^{S^{\cal J}[c,c^*,\psi]}\;,
\end{eqnarray}
with 
\begin{eqnarray}
 S^{\cal J}[c,c^*,\psi] \equiv {\int_u \int_v  c^*_u\, \frac{\delta\mbox{\small\sc{Eq}}_u[\psi]}{\delta \psi_v} \,  c_v} - \int_u c^*_u \, \frac{M''(\psi_u)}{M'(\psi_u)} \mbox{\small\sc{Eq}}_u[\psi] \, c_u \;, 
\;
\label{eq:K}
\end{eqnarray}
and the boundary conditions:
  $c(-T) = \dot c(-T) = c^*(T)  = \dot c^*(T) = 0$.
Plugging in the Langevin equation~(\ref{eq:LangevinMultiplicatif}), we arrive at
\begin{eqnarray}
 S^{\cal J}[c,c^*,\psi] &=& \int_u \int_v c^*_u \left[ m \partial_u^2 \delta_{u-v}
- \frac{\delta F_u[\psi] }{\delta \psi_v}
+ M'(\psi_u) \partial_u \gamma_{u-v} M'(\psi_v) \right] c_v \nonumber \\
& & \quad\qquad - \int_u c^*_u \, \frac{M''(\psi_u)}{M'(\psi_u)} \left[ m \partial_u^2 \psi_u -  F_u[\psi] \right]
 \, c_u \;. \label{eq:Jac_multi}
\end{eqnarray}
The Grassmann fields $c$ and $c^*$ that enter the integral representation of the determinant are known as Faddeev-Popov ghosts and can be interpreted as spinless fermions. 
The two-time fermionic Green function defined as
\begin{eqnarray}
 \langle c^*_t c_{t'}\rangle_{S^{\cal J}} &\equiv& \int \uD{[c,c^*] \, c^*_t c_{t'} \, \rme^{S^{\cal J}[c,c^*,\psi] } } \;,
\end{eqnarray}
is related, by use of Wick's theorem, to the inverse operator of $\scriptsize{\frac{\delta\mbox{\sc
{Eq}}_{t'}[\psi]}{\delta \psi_t} - \frac{M''(\psi_t)}{M'(\psi_t)}\mbox{\small\sc Eq}[\psi_t]\delta_{t-t'}}$.  $
\langle c^*_t c_{t'} \rangle_{S^{\cal J}}$ inherits the causality structure of the latter and it vanishes 
at equal times as long as the Markov limit is not taken (\textit{i.e.} all fermionic tadpole 
contributions cancel): $\langle c^*_t c_{t'} \rangle_{S^{\cal J}}= 0$ for $t\geq t'$. The last statement 
can be easily verified by considering the discretized version of $S^{\cal J}$ (see 
\ref{app:Jacobian2} and \ref{app:MSRJDmult}) and by checking that the diagonal terms of the 
inverse operator vanish in the continuous limit.
$S^{\cal J}$ only involves combinations of the form $c^* c$, \textit{i.e.} it conserves the 
fermionic charge and $\langle c_t \rangle_{S^{\cal J}}  = \langle c^*_t \rangle_{S^{\cal J}} = 0$.  
This implies that $S^{\cal J}[c,c^*,\psi]$ and more generally the MSRJD generating functional (at 
zero sources) are invariant under the following field transformation
\begin{eqnarray}
{\cal T}_{\cal J}(\alpha) \equiv 
 \left\{ \begin{array}{rcl}
          c_t &\mapsto& \alpha \, c_t \;, \\
	  c^*_t &\mapsto& \alpha^{-1} \, c^*_t \;,
         \end{array}
\right. \quad \forall\, \alpha \in  \mathbb{C}^*\;. \label{eq:TransfoGhosts}
\end{eqnarray}
The Jacobian of the transformation is trivially equal to one and the measure $\uD{[c,c^*]}$ is left 
unchanged. One has ${\cal T}_{\cal J}(\alpha) {\cal T}_{\cal J}(\beta) = 
{\cal T}_{\cal J}(\alpha\beta)$.

The total MSRJD action given in eq.~(\ref{eq:MSRJDMultip}) can be written equivalently as a 
functional of $\psi$, $\hat\psi$, $c$ and $c^*$ provided that the path-integral measure is extended 
to the newly introduced fermionic fields:
\begin{eqnarray}\label{eq:SwtFermions}
 S[\psi,\hat\psi,c,c^*] \equiv S^\mathrm{det}[\psi,\hat\psi] + S^\mathrm{diss}[\psi,\hat\psi] + 
 S^{\cal J}[c,c^*,\psi] \;.
\end{eqnarray}

\subsection{Observables}

\subsubsection{Measure.}
We denote $\langle \,...\, \rangle$ the average over the thermal noise and
the initial conditions. Within the MSRJD formalism, the average is
evaluated with respect to the action functional $S[\psi,\hat\psi]$ or  $S[\psi,\hat\psi,c,c^*]$ and
we use the notation $\langle \,...\, \rangle_{S}$:
\begin{eqnarray}
 \langle \,...\, \rangle_{S} & \equiv& \int \uD{[\psi,\hat\psi]} ... \, \rme^{S[\psi,\hat\psi]} 
=\int \uD{[\psi,\hat\psi, c, c^*]} ...\, \rme^{S[\psi,\hat\psi,c,c^*] }
\; . 
\end{eqnarray}

\subsubsection{Local observable.}
 The value of a generic local observable $A$ at time $t$ is a
function of the field and its time-derivatives evaluated at time $t$,
\textit{i.e.} a functional of the field $\psi$ around $t$,
$A([\psi],t)$. Unless otherwise specified we assume it does not depend explicitly
on time and denote it $A[\psi(t)]$. Its mean value is
\begin{eqnarray}\label{eq:defMeanA}
 \langle A[\psi(t)] \rangle
= \langle A[\psi(t)] \rangle_{S} 
\;.
\end{eqnarray}

\subsubsection{Time-reversal.}
Since it will be used in the rest of this work, we introduce the time-reversed field $\bar\psi$ by $\bar\psi(t) \equiv \psi(-t)$ for all $t$. The time-reversed observable is defined as
\begin{eqnarray}\label{eq:defAr}
A_\mathrm{r}([\psi],t) \equiv A([\bar\psi],-t).
\end{eqnarray}
It has the effect of changing the sign of all odd time-derivatives in the expression of local observables, \textit{e.g.} if $A[\psi(t)] = \partial_t \psi(t)$ then $A_\mathrm{r}[\psi(t)] = -\partial_t \psi(t)$.
As an example for non-local observables, the time-reversed Langevin equation~(\ref{eq:Langevin}) reads
\begin{eqnarray}
 \mbox{\sc{Eq}}_\mathrm{r}([\psi],t) &=& 
m \ddot \psi(t) - F_\mathrm{r}([\psi],t) - \int_{-T}^{T} 
\ud{u} \gamma(u-t)  \dot \psi(u) \;.
\end{eqnarray}
Notice the change of sign in front of the friction term that is no longer dissipative in this new equation.

\subsubsection{Two-time correlation.}
We define the two-time self correlation function as
\begin{eqnarray}
 C(t,t') &\equiv& \langle \psi(t) \psi(t') \rangle 
= \langle \psi(t) \psi(t') \rangle_{S} 
\; . 
\end{eqnarray}
Given two local observables $A$ and $B$, we similarly introduce the two-time generic correlation as
\begin{eqnarray}\label{eq:defCabCl}
 C_{\{AB\}}(t,t') 
&\equiv& 
\langle A[\psi(t)] B[\psi(t')] \rangle_{S} \;,
\end{eqnarray}
The curly brackets are here to stress the symmetry that underlies
this definition: $C_{\{AB\}}(t,t') = C_{\{BA\}}(t',t)$.

\subsubsection{Linear response.}

If we slightly modify the potential according to $ V(\psi) \mapsto
V(\psi) - f_\psi \psi$, the self linear response at time $t$ to an
infinitesimal perturbation linearly coupled to the field at a previous
time $t'$ is defined as
\begin{eqnarray} \label{eq:def_R}
 R(t,t') &\equiv&  
\left. 
\frac{\delta \langle \psi(t) \rangle}
{\delta f_\psi(t')}
\right|_{f_\psi=0}
 = 
\left. 
\frac{\delta \langle \psi(t) \rangle_{S[f_\psi]} }
{\delta f_\psi(t')}
\right|_{f_\psi=0}
\; . 
\end{eqnarray}
It is clear from causality that if $t'$ is later than $t$, $\langle
\psi(t) \rangle_{S[f_\psi]}$ cannot depend on the
perturbation $f_\psi(t')$ so $R(t,t') = 0$ for $t'>t$. At equal times,
the linear response $R(t,t)$ also vanishes as long as inertia is not
neglected ($m\neq0$)\footnote{In the double limit of a white noise and
  $m\to0$, the equal-time response can slightly violate the causality
  principle depending on the order in which the limits are taken. In the It\^o
  scenario it vanishes whereas in the Stratonovich one it has a 
  finite value.}.
More generally, the linear response of $A$ at time $t$ to an infinitesimal perturbation linearly applied to $B$ at time $t'<t$ is
\begin{eqnarray}\label{eq:defRabCl}
 R_{AB}(t,t') &\equiv&  
\left.
\frac{\delta \langle A[\psi(t)] \rangle }
{\delta f_B(t')}
\right|_{f_B=0}
=
\left.
\frac{\delta \langle A[\psi(t)] \rangle_{S[f_B]} }
{\delta f_B(t')}
\right|_{f_B=0}
\; , 
\end{eqnarray}
with $V(\psi) \mapsto V(\psi) - f_B B[\psi]$, where $B$ is local.

\subsection{Classical Kubo formula}\label{sec:langevinlike}
By computing explicitly the functional derivative $\delta/\delta f_\psi$ in the path integral generating functional, we deduce
\begin{eqnarray}
\left.\frac{\delta  \langle \, ... \, \rangle_{S[f_\psi]}}{\delta f_\psi(t)}   \right|_{f_\psi=0}  &=& \left. 
\langle \,... \, \frac{\delta S[\psi,\hat\psi,c,c^*;f_\psi] }{\delta
f_\psi(t)}
\right|_{f_\psi=0}  \hspace{-1em} \rangle_{S} 
\nonumber \\
&=&
\langle \,... \, \rmi\hat\psi(t)  \rangle_S 
+
 \langle \, ... \, \frac{M''(\psi(t))}{M'(\psi(t))  } c^*(t)c(t) \rangle_{S}\;.
\end{eqnarray}
The first term in the {\sc rhs} comes from the functional derivative of $S^\mathrm{det}$. The second
term comes from the Jacobian term expressed with the fermionic ghosts, $S^{\cal J}$, and 
vanishes identically (see the discussion on the equal-time fermionic Green function in 
Sect.~\ref{sec:Jaco}). One has
\begin{eqnarray}
 && \langle \rmi\hat\psi(t) \rangle_{S} 
=  
\left.  \frac{\delta \langle \, 1  \,\rangle_{S[f_\psi]} }
{\delta f_\psi(t)} \right|_{f_\psi=0} = 0 \;, \\
 &&
 \langle \rmi\hat\psi(t) \rmi\hat\psi(t')\rangle_{S} 
= 
\left.  \frac{\delta^2 \langle \, 1 \, \rangle_{S[f_\psi]} }
{\delta f_\psi(t) \ \delta f_\psi(t')} \right|_{f_\psi=0} = 0 
\;. \label{eq:hatpsihatpsi}
\end{eqnarray}
From the definition of the linear response, eq.~(\ref{eq:def_R}), we
deduce the `classical Kubo formula'~\cite{Kubo}
\begin{eqnarray}
 R(t,t') =  \langle \psi(t) \rmi\hat\psi(t') \rangle_{S}\;.
\label{eq:response-psihat}
\end{eqnarray}
The linear response is here written within the MSRJD formalism as a correlation computed with an unperturbed
action. The causality of the response is not explicit, nevertheless following the lines in
Ref.~\cite{Lubensky} one can check it is built-in\footnote{In general, a multi-time correlator involving $\rmi\hat\psi(t_1)$ vanishes if $t_1$ is the largest time involved.}. Because of this expression, the auxiliary field $\hat\psi$ is often called
the response field. Observe that we have not specified the nature of the
initial probability distribution $P_\mathrm{i}$ nor the driving forces; eq.~(\ref{eq:response-psihat})  
holds even out of equilibrium.

Similarly, by plugging  eq.~(\ref{eq:defMeanA})
into eq.~(\ref{eq:defRabCl}),
we obtain the classical Kubo formula for generic local observables:
\begin{eqnarray} \label{eq:RabCl}
 R_{AB}(t,t') &=& \langle A[\psi(t)] 
\left.
\frac{\delta S[\psi,\hat\psi,c,c^*;f_B] }{\delta f_B(t')}
\right|_{f_B=0} \hspace{-1em} \rangle_{S} 
\nonumber\\
&=& \langle A[\psi(t)] \int \ud{u} \rmi\hat\psi(u)\frac{\delta B[\psi(t')] }{\delta \psi(u)} 
\rangle_{S} \nonumber\\
&=& \langle A[\psi(t)] \sum_{n=0}^{\infty} \partial^n_{t'} \rmi\hat\psi(t') \
\frac{\partial B[\psi(t')] }{\partial\ \partial^n_{t'} \psi(t')} 
\rangle_{S} \;.
\end{eqnarray}
This formula is valid in and out of equilibrium and allows us to write the response functions associated 
to generic observables (\textit{e.g.} functions of the position, velocity, acceleration, kinetic energy, etc.)
as correlators of $\psi$, $\hat\psi$ and their time derivatives. For example if $B$ is just a function of the field (and not of its time-derivatives), only the $n=0$-term subsists in the above sum, yielding
\begin{eqnarray} 
 R_{AB}(t,t') &=&  \langle A[\psi(t)] \rmi\hat\psi(t') \frac{\partial
B[\psi(t')] }{\partial \psi(t')} 
\rangle_{S} \;.
\end{eqnarray}
As another example, if one is interested in the response of the acceleration
$A[\psi(t)] =\partial^2_t\psi(t)$ to a perturbation of the kinetic energy
$B[\psi(t)]=\frac{1}{2}m(\partial_t\psi(t))^2$ one should compute
\begin{eqnarray} 
 R_{AB}(t,t') &=& m \langle \partial^2_t\psi(t) \partial_{t'} \rmi\hat\psi(t')
\partial_{t'} \psi(t')\rangle_{S} \;.
\end{eqnarray}
Furthermore, it is straightforward to see that within the MSRJD formalism we can
extend all the previous definitions and formul{\ae} to $A$ being a local functional
of the auxiliary field: $A[\hat\psi(t)]$. For example, if
$A[\hat\psi(t)]=\rmi\hat\psi(t)$ and $B[\psi(t)] = \psi(t)$, we obtain the mixed
response
\begin{eqnarray} 
 R_{\rmi\hat\psi\psi}(t,t') &=&  \langle \rmi\hat \psi(t) \rmi\hat\psi(t')
\rangle_{S} = 0\;, \label{eq:Rhat}
\end{eqnarray}
where we used eq.~(\ref{eq:hatpsihatpsi}).

\section{Equilibrium}
\label{sec:equilibrium}

In this Section we focus on situations in which the system is in equilibrium.
We identify a field transformation that leaves the MSRJD generating functional (evaluated
at zero sources) invariant. The corresponding Ward-Takahashi identities between 
the expectation values of different observables imply a number of model independent 
equilibrium properties including stationarity, Onsager relations and the fluctuation-dissipation theorem (FDT). 
These proofs are straightforward in the generating functional formalism, demonstrating
its advantage  with respect to the Fokker-Planck or master equation ones, 
when the environment acts multiplicatively and has a non-vanishing correlation time.
We shall report soon~\cite{Arbicu} on
the extension to the quantum case where the Keldysh action also exhibits a
non-trivial symmetry for equilibrium dynamics. Similarly to the
classical case, this symmetry leads to the quantum FDT.

\subsection{The action}

Equilibrium dynamics are guaranteed provided that, apart from its interactions
with the bath, the system is prepared and subjected to the same
time-independent and conservative forces ($F=-V'$). In such
situations, the initial state is taken from the Boltzmann probability
distribution
\begin{equation}
\ln P_\mathrm{i}(\psi_{-T},\dot\psi_{-T}) = 
- \beta {\cal H}[\psi_{-T}]
- \ln {\cal Z}
\; , 
\end{equation}
where ${\cal H}[\psi_t] \equiv \frac{1}{2}m \dot\psi_t^2 + V(\psi_t)$ is the internal energy of the system, and
${\cal Z}$ is the partition function.
The Langevin evolution of the system in contact with the bath can be put in the form
\begin{equation}
-\int_u \frac{\delta {\cal L}[\psi_u]}{\delta \psi_t}
 + M'(\psi_t) \int_u
 \gamma_{t-u} 
M'(\psi_u)
 \dot \psi_u
=
M'(\psi_t) \xi_t \;,
\end{equation}
with ${\cal L}[\psi_u] \equiv \frac{1}{2}m \dot\psi_u^2 - V(\psi_u)$
being the Lagrangian of the system.  In this equilibrium set-up, the
deterministic part of the MSRJD action functional reads
\begin{eqnarray}
S^\mathrm{det}[\psi, \hat\psi] 
=
-\beta {\cal H}[\psi_{-T}] - \ln {\cal Z} + \int_u\int_v  \rmi\hat \psi_u \frac{\delta {\cal L}[\psi_v]}{\delta \psi_u}    
\nonumber\\
\quad \;\; =
-\beta \left(\frac{1}{2}m\dot\psi^2_{-T} + V(\psi_{-T}) \right)- \ln {\cal Z}
 - \int_u \rmi\hat \psi_u \left[ m\ddot \psi_u + 
 V'(\psi_u)  \right] . \label{eq:Sdet_eq}
\end{eqnarray}
The dissipative part of the MSRJD action functional remains the same, see
eq.~({\ref{eq:Sdiss_multi}}). 
As discussed in Sect.~\ref{sec:Jaco}, the Jacobian ${\cal J}_0$ enters the action through the constant term $\ln {\cal J}_0$ or it can be expressed in terms of a Gaussian integral over the ghosts fields $c$ and $c^*$. In this case, its contribution to the action reads
\begin{eqnarray}
&&
S^{\cal J}[c,c^*,\psi] = \int_u \int_v c^*_u \left[ m \partial_u^2 \delta_{u-v}
+ M'(\psi_u) \partial_u \gamma_{u-v} M'(\psi_v) \right] c_v
\nonumber \\
&& 
\qquad \;\; -
\int_u c^*_u\left[ -V''(\psi_u)+ \frac{M''(\psi_u)}{M'(\psi_u)} \partial_u^2 \psi_u
 +\frac{M''(\psi_u)}{M'(\psi_u)} V'(\psi_u) \right] c_u 
 \;.  
 \label{eq:Jac_eq}
\end{eqnarray}

\subsection{Symmetry of the MSRJD generating functional}\label{sec:FDTSym}

We shall prove that $\int \uD{[\psi,\hat\psi,c,c^*]} \rme^{S[\psi,\hat\psi,c,c^*]}
\nonumber$ is invariant under the equilibrium field transformation:
\begin{eqnarray}\label{eq:TransfoFDT}
{\cal T}_\mathrm{eq} \equiv
\left\{ 
 \begin{array}{rclrcl}
 \psi_u &\mapsto& \psi_{-u}\;,& \qquad  c_u &\mapsto&  c^*_{-u}\;, \\
 \rmi\hat\psi_u &\mapsto&\rmi\hat\psi_{-u} + \beta\partial_u \psi_{-u}\;,& \qquad c^*_u &\mapsto& - c_{-u} \;.
 \end{array}
\right.
\end{eqnarray}
This transformation is involutary, $\mathcal{T}_\mathrm{eq}
\mathcal{T}_\mathrm{eq} = 1$, when applied to the fields $\psi$ or $\rmi\hat \psi$ and the composite field $c^*c$.
 It does not involve the kernel
$\gamma$ and includes a time-reversal. It is interesting to reckon
that the invariance is achieved independently by the deterministic
($S^\mathrm{det}$), the dissipative ($S^\mathrm{diss}$) and the Jacobian
($S^{\cal J}$) contributions to the action. This means that it is
still valid in the Newtonian limit ($\gamma=0$).  The detailed proof
that we develop here consists of two parts: we first show that the
Jacobian of the transformation is unity, then that the integration
domain of the transformed fields is unchanged. Afterwards we show that
the action functional $S[\psi,\hat\psi, c, c^*]$ is invariant under
$\mathcal{T}_\mathrm{eq}$.

\subsubsection{Invariance of the measure.}\label{app:SymC1}

The equilibrium transformation  $\mathcal{T}_\mathrm{eq}$ acts 
separately on the fields $\psi$ and $\rmi \hat\psi$ on the
one hand, and the fields $c$ and $c^*$ on the other. 
The Jacobian $\mathcal{J}_\mathrm{eq}$ thus factorizes into a bosonic part and a fermionic part.
The bosonic part is the determinant of a
triangular matrix:
\begin{eqnarray}
 \mathcal{J}^\mathrm{b}_\mathrm{eq} &\equiv& \mbox{det} \left[ \frac{\delta (\psi,\hat\psi )}{\delta ({\cal T}_\mathrm{eq} \psi, {\cal T}_\mathrm{eq} \hat\psi)} \right]
= \mbox{det}_{uv}^{-1}
	\left[ 		\begin{array}{cc}
			 \frac{\delta \psi_{-u}}{\psi_{v}} & 0 \\
			 \frac{\delta \hat\psi_{-u}}{\psi_{v}} &  \frac{\delta \hat\psi_{-u}}{\hat\psi_{v}}
			\end{array}
	\right] 
  =   \left( \mbox{det}^{-1}_{uv}\left[\delta_{u+v}\right]\right)^2 = 1\;,  \nonumber
\end{eqnarray}
and it is thus identical to one~\cite{Velenich}.  It is easy to verify that the fermionic part
$\mathcal{J}^\mathrm{f}_\mathrm{eq}=1$ as well.

\subsubsection{Invariance of the integration domain.}\label{app:SymC2}

Before and after the transformation, the functional integration on the
field $\psi$ is performed for values of $\psi_t$ on the real
axis. However, the new domain of integration for the field $\hat\psi$
is complex. For a given time $t$, $\hat\psi_t$ is now integrated over the
complex line with a constant imaginary part
$-\rmi\beta\partial_t\psi_t$. One can return to an integration
over the real axis by closing the contour at both infinities. Indeed,
the integrand, $\rme^{S}$, goes to zero sufficiently fast at
$\psi_t\to\pm\infty$ for neglecting the vertical ends of the contour
thanks to the term $\beta^{-1}\gamma_0(\rmi\hat\psi_t)^2 $ in the
action. Furthermore, the new field is also integrated with the boundary 
conditions $\hat\psi(-T) = \hat\psi(T) =0$.

The equilibrium transformation 
leaves the measure $\uD{[c,c^*]}$ unchanged
together with the set of boundary conditions $c(-T) = \dot
c(-T) = c^*(T) = \dot c^*(T) = 0$.

\subsubsection{Invariance of the action functional.}\label{app:SymC3}
The MSRJD action functional 
$S[\psi,\hat\psi,c,c^*] = S^\mathrm{det}[\psi, \hat\psi] +S^\mathrm{diss}[\psi, \hat\psi] +  S^{\cal J}(c,c^*,\psi)$
is invariant term by term. The deterministic contribution given in 
eq.~(\ref{eq:Sdet_eq}) satisfies 
\begin{eqnarray}
&& 
S^\mathrm{det}[\mathcal{T}_\mathrm{eq} \psi, \mathcal{T}_\mathrm{eq}\hat\psi] 
= 
\ln P_\mathrm{i}(\psi_T, \dot\psi_T) 
- \int_u 
[ \rmi\hat \psi_{-u}  + \beta \partial_u \psi_{-u}]
[ m \partial^2_u  \psi_{-u} + V'(\psi_{-u}) ]
\nonumber\\
&& \qquad = 
\ln P_\mathrm{i}(\psi_T, \dot\psi_T) 
-\int_u 
\rmi\hat \psi_u [ m\ddot \psi_u + V'(\psi_u)  ]
+ \beta \int_u 
\dot\psi_u [ m \ddot\psi_u + V'(\psi_u)  ] 
\nonumber\\
&& \qquad =
\ln P_\mathrm{i}(\psi_T, \dot\psi_T) 
- \int_u
\rmi\hat \psi_u [ m\ddot\psi_u + V'(\psi_u)  ]
+
\beta \int_u \partial_u {\cal H}[\psi_u]
\nonumber\\
&& \qquad = 
S^\mathrm{det}[\psi, \hat\psi] 
\;, 
\end{eqnarray}
where we used the initial equilibrium measure $\ln P_{\rm
  i}(\psi,\dot\psi) = -\beta {\cal H}[\psi] - \ln{\cal Z}$. In the first line we just applied the
transformation, in the second line we made the substitution $u\mapsto
-u$, in the third line we wrote the last integrand as a total
derivative the integral of which cancels the first term and creates a
new initial measure.

Secondly, we show that the dissipative contribution $S^{\rm
  diss}[\psi, \hat\psi]$, defined in eq.~(\ref{eq:Sdiss}), is also
invariant under the equilibrium transformation.
We have
\begin{eqnarray}
S^\mathrm{diss}[\mathcal{T}_\mathrm{eq}\psi,\mathcal{T}_\mathrm{eq}\hat\psi] 
&=&
\int_u 
[ \rmi\hat \psi_{-u} + 
\beta \partial_u \psi_{-u}] 
\int_v \beta^{-1} M'(\psi_{-u}) \, \gamma_{u-v}\,  M'(\psi_{-v}) \  \rmi\hat \psi_{-v} 
\nonumber\\
&=&
\int_u [ \rmi\hat \psi_u - \beta \dot\psi_u] 
\int_v M'(\psi_{u}) \, \gamma_{v-u} \, M'(\psi_{v}) \beta^{-1} \rmi\hat \psi_v   
\nonumber\\
&=& S^\mathrm{diss}[\psi, \hat\psi]
\; . 
\end{eqnarray}
In the first line we just applied the transformation, in the second
line we made the substitution $u\mapsto -u$ and in the last step we
exchanged $u$ and $v$.

Finally, we show that the Jacobian term in the action is invariant once it is expressed in terms of a
Gaussian integral over conjugate Grassmann fields ($c$ and $c^*$).
We start from eq.~(\ref{eq:Jac_eq})
\begin{eqnarray}
&&
 S^{\cal J}[{\cal T}_\mathrm{eq} c, {\cal T}_\mathrm{eq} c^*, {\cal T}_\mathrm{eq} \psi]
\nonumber\\
&&
\qquad
=
-\int_u \int_v c_{-u} \left[ m \partial_u^2 \delta_{u-v}
+ M'(\psi_{-u}) \partial_u \gamma_{u-v} M'(\psi_{-v}) \right] c^*_{-v}
\nonumber \\
&&
\qquad\;\;\;\;
+ \int_u c_{-u}\left[ -V''(\psi_{-u})+ \frac{M''(\psi_{-u})}{M'(\psi_{-u})} \partial_u^2 \psi_{-u}
 +\frac{M''(\psi_{-u})}{M'(\psi_{-u})} V'(\psi_{-u}) \right] c^*_{-u} 
 \nonumber \\
&&
\qquad
=
\int_u \int_v c^*_{v} \left[ m \partial_u^2 \delta_{v-u}
- M'(\psi_{u}) \partial_u \gamma_{v-u} M'(\psi_{v}) \right] c_{u}
\nonumber \\
&& 
\qquad\;\;\;\; 
 -\int_u c^*_{u}\left[ -V''(\psi_{u})+ \frac{M''(\psi_{u})}{M'(\psi_{u})} \partial_u^2 \psi_{u}
 +\frac{M''(\psi_{u})}{M'(\psi_{u})} V'(\psi_{u}) \right] c_{u} \nonumber \\
&&
\qquad
= S^{\cal J}[c,  c^*, \psi]
\;.  
\end{eqnarray}
In the first line we just applied the transformation, in the second
line we exchanged the anti-commuting Grassmann variables and made the
substitutions $u\mapsto-u$ and $v\mapsto-v$, in the last step we used
$\partial_v \gamma_{v-u} = - \partial_v \gamma_{u-v}$ and exchanged
$u$ and $v$.

\subsection{Ward-Takahashi identities}
We just proved that equilibrium dynamics manifest themselves as a symmetry of the MSRJD action and more generally at the level of the generating functional. This symmetry has direct consequences at the level of correlation functions. If $A$ is a generic functional of $\psi$ and $\hat\psi$, it implies
 the following Ward-Takahashi identity
\begin{eqnarray}
 \langle A[\psi,\hat\psi] \, ... \, \rangle_{S} 
= \langle A[\mathcal{T}_\mathrm{eq}\psi,\mathcal{T}_\mathrm{eq}\hat\psi] \, ... \,  \rangle_{S} \;.
\end{eqnarray}
This identity leads to all the possible equilibrium relations between observables as we shall now describe in the following. These relations can be proven without using the MSRJD path integral formalism, however 
our point is to show that the symmetry is able to generate all the equilibrium relations without using any other ingredient.

\subsection{Stationarity}

In equilibrium, one expects noise-averaged observables to be independent of the time $t_0$ at which the system was prepared (in our case $t_0 = -T$). One-time dependent noise-averaged observables are expected to be constant, $\langle A[\psi_t] \rangle  = \mathrm{ct}$, and two-time correlations to be time-translational invariant:
$\langle A[\psi_t] B[\psi_{t'}] \rangle = f_{t-t'}$. Similarly, one argues that
multi-time correlations can only depend upon all possible independent
time-differences between the times involved. These statements have been proven 
for additive white noise processes using the Fokker-Planck~\cite{Montanari} formalism.
The use of the transformation ${\cal T}_\mathrm{eq}$ allows one to show
these properties very easily for generic Langevin processes.

\paragraph{One-time observables.}
Taking $A = 1$ and letting $B$ be a generic local observable,
 the equal-time linear response vanishes, $R_{AB}(t,t) = 0$.
Using the classical Kubo formula~(\ref{eq:RabCl}), 
\begin{eqnarray}
  {R_{AB}}(t,t) = \langle  \sum_{n=0}^{\infty} \partial^n_{t} \rmi\hat\psi_t \
\frac{\partial B[\psi_t] }{\partial\ \partial^n_{t} \psi_t} 
\rangle_{S}  = 0
\;.
\end{eqnarray}
Applying the transformation ${\cal T}_\mathrm{eq}$, we find
\begin{eqnarray}
  {R_{AB}}(t,t)
 &=& 
\langle  \sum_{n=0}^{\infty} \partial^n_{t} \rmi\hat\psi_{-t} \
\frac{\partial B_\mathrm{r}[\psi_{-t}] }{\partial\ \partial^n_{t} \psi_{-t}} 
\rangle_{S} 
+
\beta \langle \sum_{n=0}^{\infty} \partial^{n+1}_{t} \psi_{-t} \
\frac{\partial B_\mathrm{r}[\psi_{-t}] }{\partial\ \partial^n_{t} \psi_{-t}} 
\rangle_{S}\;.
\end{eqnarray}
The {\sc lhs} and the first term in the {\sc rhs} vanish
identically at all times. One is left with the second term in the {\sc rhs} that simply reads
$\langle \partial_t B_\mathrm{r}[\psi_{-t}] \rangle = 
\partial_t \langle B_\mathrm{r}[\psi_{-t}] \rangle = 0$, proving 
that all one-time local observables are constant in time.

\paragraph{Two-time observables.}
Because we just showed that $\langle A[\psi(t)] \rangle$ is constant in equilibrium, the response 
$R_{AB}(t,t')$, see its formal definition in eq.~(\ref{eq:defRabCl}), can only be a function of the 
time-difference between the observation time and the time at which the perturbation is applied. 
Therefore it can be written in the form $R_{AB}(t,t') = f(t-t')\theta(t-t')$. We shall see in Sect.~\ref
{sec:FDT} that the fluctuation-dissipation theorem relates, in equilibrium, the linear response 
$R_{AB}(t,t')$ to the two-time correlation $C_{\{AB\}}(t,t')$ implying that this last quantity is also 
time-translational invariant.

Similarly, $(n+1)$-time correlators can be proven to be functions of $n$ independent time-differences because they are related, in equilibrium, to responses of $n$-time correlators that are 
time-translational invariant.

\subsection{Equipartition theorem}

Let us consider the local observables $A[\psi(t)]=\partial_t \psi(t)$ and $B[\psi(t)]=\psi(t)$. 
The linear response is ${R_{AB}}(t,t') = \langle \partial_t \psi_t \rmi\hat\psi_{t'} \rangle_S = \partial_t 
\langle  \psi_t \rmi\hat\psi_{t'} \rangle_S $  and we recognize $ \partial_t R(t,t')$. Using the field 
transformation ${\cal T}_\mathrm{eq}$, we find
\begin{eqnarray}
  \partial_t R(t,t') &=&  \partial_{t} \langle  \psi_{-t} \rmi\hat\psi_{-t'} \rangle_S
 + \beta  \langle  \partial_{t} \psi_{-t}  \partial_{t'} \psi_{-t'} \rangle_S 
 \nonumber \\
&=&  \partial_{t} \langle  \psi_{-t} \rmi\hat\psi_{-t'} \rangle_S
 + \beta  \langle  \partial_{t} \psi_{t}  \partial_{t'} \psi_{t'} \rangle_S 
 \; . 
\end{eqnarray}
If $t>t'$, the first term in the {\sc rhs} vanishes by causality. Considering moreover the limit $t'\to t^-$ the {\sc lhs} is  $1/m$ as we shall show in Sect.~\ref{sec:o-e-relations}. Finally, we recover the equipartition theorem for the kinetic energy
\begin{eqnarray}
 \beta m \langle \left( \partial_t \psi_t \right)^2 \rangle = 1\;.
\end{eqnarray}

\subsection{Reciprocity relations}

If we use ${\cal T}_\mathrm{eq}$ in the expression~(\ref{eq:defCabCl}) of
generic two-time correlation functions, we have
\begin{eqnarray}
 \langle A[\psi_t] B[\psi_{t'}] \rangle_{S} = \langle A_\mathrm{r}[\psi_{-t}] B_\mathrm{r}[\psi_{-t'}] \rangle_{S}
 \;,
\end{eqnarray}
reading
\begin{eqnarray}
C_{\{AB\}}(t,t') = C_{\{A_\mathrm{r} B_\mathrm{r}\}}(-t,-t')\;.
\end{eqnarray}
In cases in which $A$ and $B$ have a definite parity under time-reversal:
\begin{eqnarray}
 C_{\{AB\}}(\tau) &=& C_{\{A B\}}(|\tau|) \mbox{ if } A \mbox{ and } B \mbox{ have the same parity,} \nonumber \\
 C_{\{AB\}}(\tau) &=& -C_{\{A B\}}(-\tau) \mbox{ otherwise.} \nonumber
\end{eqnarray}

\subsection{Fluctuation-dissipation theorem (FDT)}\label{sec:FDT}

\subsubsection{Self FDT.}

Applying the transformation to the
expression~(\ref{eq:response-psihat}) of the self response $R(t,t')$
we find
\begin{eqnarray}
\langle \psi_t \rmi\hat \psi_{t'} \rangle_{S} &=& 
\langle \mathcal{T}_\mathrm{eq} \psi_t \mathcal{T}_\mathrm{eq}
\rmi\hat \psi_{t'} \rangle_{S} 
= \langle \psi_{-t} \rmi\hat \psi_{-t'} \rangle_{S}   + 
\beta \langle \psi_{-t} \partial_{t'} \psi_{-t'} \rangle_{S}
\; , 
\end{eqnarray}
and we read
\begin{eqnarray}
R(t,t') &=& R(-t,-t') +  \beta \partial_{t'} C(-t,-t')\; 
\end{eqnarray}
that, using the equilibrium time-translational invariance, becomes 
\begin{eqnarray}
R(\tau)-R(-\tau) &=& - \beta \partial_{\tau} C(-\tau)
\;, 
\end{eqnarray}
where we set $\tau \equiv t-t'$.
Since $C(\tau)$ is symmetric in $\tau$ by definition, this expression can be 
rewritten, once multiplied by $\Theta(\tau)$, as
\begin{eqnarray}
 R(\tau) &=& - \Theta(\tau) \beta \partial_\tau C(\tau)\;. 
 \label{eq:FDT1}
\end{eqnarray}
Equation~(\ref{eq:FDT1}) is the well-known fluctuation-dissipation
theorem. It allows one to predict the slightly out of equilibrium behavior
of a system -- such as the irreversible dissipation of energy into
heat -- from its reversible fluctuations in equilibrium.  

\subsubsection{Generic two-time FDTs.}

We generalize the previous FDT relation to the case of generic local
observables $A$ and $B$.  Applying the equilibrium transformation
$\mathcal{T}_\mathrm{eq}$ to expression~(\ref{eq:RabCl}) of the linear
response $R_{AB}(t,t')$
\begin{eqnarray}
 \fl \qquad \langle A[\psi_t] \sum_{n=0}^{\infty} \partial^{n}_{t'} 
 \rmi\hat\psi_{t'} \ \frac{\partial B[\psi_{t'}]}{\partial \ \partial^{n}_{t'} \psi_{t'}}  \rangle_{S} 
=
\langle A_\mathrm{r}[\psi_{-t}] \sum_{n=0}^{\infty} \partial^{n}_{t'} \rmi\hat\psi_{-t'} \ 
\frac{\partial B_\mathrm{r}[\psi_{-t'}]}{\partial \ \partial^{n}_{t'} \psi_{t'}}  \rangle_{S}  \nonumber 
\\ \qquad\qquad\qquad\qquad\qquad + 
\beta \ \langle A_\mathrm{r}[\psi_{-t}] 
\sum_{n=0}^{\infty} \partial^{n+1}_{t'} \psi_{-t'} \ 
\frac{\partial B_\mathrm{r}[\psi_{-t'}]}{\partial \ \partial^{n}_{t'} \psi_{t'}}
\rangle_{S} 
\nonumber\\
= 
\langle A_\mathrm{r}[\psi_{-t}] \sum_{n=0}^{\infty} \partial^{n}_{t'} \rmi\hat\psi_{-t'} \ 
\frac{\partial B_\mathrm{r}[\psi_{-t'}]}{\partial \ \partial^{n}_{t'} \psi_{t'}}  \rangle_{S} \nonumber 
+ \beta \ \partial_{t'} \langle A_\mathrm{r}[\psi_{-t}]B_\mathrm{r}[\psi_{-t'}] \rangle_{S} 
\;.
\end{eqnarray}
Applying once again the transformation to the last term in the {\sc{rhs}} yields
\begin{eqnarray}
 \fl \langle A[\psi_t] \sum_{n=0}^{\infty} \partial^{n}_{t'} 
 \rmi\hat\psi_{t'} \ \frac{\partial B[\psi_{t'}]}{\partial \ \partial^{n}_{t'} \psi_{t'}}  \rangle_{S} 
&=& \langle A_\mathrm{r}[\psi_{-t}] \sum_{n=0}^{\infty} \partial^{n}_{t'} 
\rmi\hat\psi_{-t'} \ \frac{\partial B_\mathrm{r}[\psi_{-t'}]}{\partial \ \partial^{n}_{t'} \psi_{t'}}  \rangle_{S} 
+ \beta \partial_{t'} \langle A[\psi_{t}]B[\psi_{t'}] \rangle_{S},
\end{eqnarray}
which reads
\begin{eqnarray}\label{eq:Rab-Rarbr}
 R_{AB}(\tau) - R_{A_\mathrm{r} B_\mathrm{r}}(-\tau) &=& -\beta \partial_{\tau} C_{\{A B\}}(\tau)
\;.
\end{eqnarray}
By multiplying both sides by $\Theta(\tau)$ we obtain the FDT for any local $A$ and $B$
\begin{eqnarray}\label{eq:classicalFDT}
 R_{AB}(\tau) &=& -\Theta(\tau) \beta \partial_\tau C_{\{AB\}}(\tau)
\; . 
\end{eqnarray}

\subsection{Higher-order FDTs: e.g. 3-time observables}

We give a derivation, \textit{via} the symmetry of the MSRJD
formalism, of relations shown and discussed in, {\it e.g.}~\cite{Montanari},
within the Fokker-Planck formalism for stochastic processes with white noise.

\subsubsection{Response of a two-time correlation.}
We first look at the response of a two-time correlator to a linear perturbation applied at time $t_1$
\begin{eqnarray}
 R(t_3,t_2;t_1) \equiv  \left. \frac{\delta \langle \psi_{t_3} \psi_{t_2} \rangle}{\delta f_{{\psi}_{t_1}}} \right|_{f_\psi=0}\;.
\end{eqnarray}
In the MSRJD formalism, it can be expressed as the 3-time correlator
\begin{eqnarray}
 R(t_3,t_2;t_1) =  \langle \psi_{t_3} \psi_{t_2} \rmi\hat\psi_{t_1} \rangle_{S}
\;.
\end{eqnarray}
Causality ensures that the response vanishes if the perturbation is posterior to
the observation times: $R(t_3,t_2;t_1) = 0$  if $t_1 > \max(t_2,t_3)$. We assume without loss of generality that $t_2 < t_3$. Under equilibrium
conditions, the response transforms under $\mathcal{T}_\mathrm{eq}$ as 
\begin{eqnarray}
 R(t_3,t_2;t_1) =  \langle \psi_{-t_3} \psi_{-t_2} \rmi\hat\psi_{-t_1} \rangle_{S}
 + \beta \partial_{t_1} \langle \psi_{-t_3} \psi_{-t_2} \psi_{-t_1} \rangle_{S}
 \;.
\end{eqnarray}
Multiplying both sides by $\Theta(t_3-t_1) $ and transforming
once again the last term in the {\sc rhs}, the last equation can be written in the 
form
\begin{equation}
 R(t_3,t_2;t_1) = \left\{ \hspace{-1ex}
\begin{array}{l}
\beta \partial_{t_1} \langle \psi_{t_3} \psi_{t_2} \psi_{t_1} \rangle_{S}  \mbox{  if } t_1 < t_2 < t_3 \;, \\ 
 R(-t_3,-t_2;-t_1) + \beta \partial_{t_1} \langle \psi_{t_3} \psi_{t_2} \psi_{t_1} \rangle_{S}  \mbox{  if } t_2 <  t_1 < t_3 \,, \\
0 \ \mbox{  if } t_2 <  t_3 < t_1\;.
\end{array}
\right. \label{eq:3FDT}
\end{equation}

\subsubsection{Second order response.}
Let us now look at the response to a perturbation at time $t_1$ of the linear response $R(t_3,t_2)$:
\begin{eqnarray}
 R(t_3;t_2,t_1) \equiv  \left. \frac{\delta^2 \langle \psi_{t_3} \rangle}{\delta f_{{\psi}_{t_1}} \ \delta f_{{\psi}_{t_2}}} \right|_{f_\psi=0}\;.
\end{eqnarray}
In the MSRJD formalism, it can be expressed as the 3-time correlator
\begin{eqnarray}
 R(t_3;t_2,t_1) =  \langle \psi_{t_3} \rmi\hat\psi_{t_2} \rmi\hat\psi_{t_1} \rangle_{S}
\;.
\end{eqnarray}
It is clear from causality that the response vanishes if the observation time is before the two 
perturbations: $R(t_3;t_2,t_1) = 0$ if $ t_3<\min(t_1,t_2)$. The response transforms under $
\mathcal{T}_\mathrm{eq}$ as 
\begin{eqnarray}
R(t_3;t_2,t_1) &=&  R(-t_3;-t_2,-t_1)
+ \beta \partial_{t_1} R(-t_3,-t_1;-t_2)  \nonumber \\
&+& \beta \partial_{t_2} R(-t_3,-t_2;-t_1) 
+ \beta^2 \partial_{t_1} \partial_{t_2} \langle \psi_{-t_3} \psi_{-t_2} \psi_{-t_1} \rangle_{S} 
\; .
\end{eqnarray}
Let us assume without loss of generality that $t_1 < t_2$. Using causality arguments and applying 
once more the equilibrium transformation to the remaining terms, 
\begin{eqnarray}
 R(t_3;t_2,t_1) = \left\{
\begin{array}{l}
0 \ \mbox{ if } t_3 < t_1 < t_2 \;,  \\
+ \beta \partial_{t_1} R(t_3,t_1;t_2)
 \ \mbox{ if } t_1 < t_3 < t_2 \;, \\
\beta \partial_{t_1} R(t_3,t_1;t_2) \ \mbox{ if }t_1 < t_2 < t_3 \;.
\end{array}
\right.
\end{eqnarray}

\subsection{Onsager reciprocal relations}
Rewriting twice eq.~(\ref{eq:Rab-Rarbr}) as
\begin{eqnarray}
 R_{AB}(\tau) - R_{A_\mathrm{r}B_\mathrm{r}}(-\tau) &=& -\beta \partial_{\tau} C_{\{ AB\}}(\tau) \;, \\
 R_{BA}(-\tau) - R_{B_\mathrm{r}A_\mathrm{r}}(\tau) &=& \beta \partial_{\tau} C_{\{ BA\}}(-\tau) = \beta \partial_{\tau} C_{\{ AB\}}(\tau) \;,
\end{eqnarray}
and summing up these two equations with $\tau>0$
\begin{eqnarray}
 R_{AB}(\tau) &=& R_{B_\mathrm{r}A_\mathrm{r}}(\tau)  \;.
\end{eqnarray}
These equilibrium relations, known as the Onsager reciprocal relations, express the fact that the linear response of an observable $A$ to a perturbation coupled to another observable $B$ can be deduced by the response of $B_\mathrm{r}$ to a perturbation coupled to $A_\mathrm{r}$.

\subsection{Supersymmetric formalism} \label{sec:susy}
\subsubsection{Generating functional.}

The generating functional of stochastic equations with conservative
forces admits a supersymmetric formulation. This has been derived and
discussed for additive noise in a number of
publications~\cite{ParisiSourlas, Zinn-Justin,Jorge}. We extend it here to
multiplicative non-Markov Langevin processes (see~\cite{Barci} for 
a study of the massless and white noise limits). To this end, let us
introduce $\theta$ and $\theta^*$, two anticommuting Grassmann
coordinates, and the superfield
\begin{eqnarray}
 \Psi(t,\theta,\theta^*) \equiv \psi(t) + c^*(t)\,  \theta + \theta^*\,  c(t) + \theta^*\theta \, \left( \rmi\hat\psi
 (t) + c^*(t)\, c(t) \, \frac{M''(\psi(t))}{M'(\psi(t))} \right) \;. \nonumber
\end{eqnarray}
The MSRJD action $S$ [see eq.~(\ref{eq:SwtFermions})] has a compact representation in terms of 
this superfield:
\begin{eqnarray}\label{eq:SPsi}
 S = S_\mathrm{susy}^\mathrm{det} + S_\mathrm{susy}^\mathrm{diss} \;,
\end{eqnarray}
with
\begin{eqnarray} 
 \hspace{-0.5cm} S_\mathrm{susy}^\mathrm{det}[\Psi] \equiv - \beta \int \Udd{\theta}{\theta^*} \theta^*\theta \, 
 {\cal H}[\Psi(-T, \theta,\theta^*)] - \ln {\cal Z} +  \int \Ud{t} \udd{\theta}{\theta^*} {\cal L}[\Psi]\;, \\
\hspace{-0.5cm} S_\mathrm{susy}^\mathrm{diss}[\Psi] \equiv \frac{1}{2} \iint \Udd{\Upsilon'}{\Upsilon} M(\Psi
(\Upsilon'))\,  \mathbf{D}^{(2)}(\Upsilon', \Upsilon) \, M(\Psi(\Upsilon))\;,
\end{eqnarray}
${\cal H}[\Psi] \equiv \frac{1}{2} m \dot\Psi^2 + V(\Psi)$ and ${\cal L}[\Psi] \equiv \frac{1}{2} m \dot
\Psi^2 - V(\Psi)$. In the second equation above we used the notation 
 $\Upsilon \equiv (t,\theta, \theta^*)$ and 
 $\udm{\Upsilon} \equiv \ud{t}\ud{\theta}\udm{\theta^*}$. The `dissipative' differential operator is 
 defined as
\begin{eqnarray}
  \mathbf{D}^{(2)}(\Upsilon', \Upsilon) &\equiv& \gamma(t'-t) \delta({\theta^*}'-\theta^*) \delta(\theta'-
 \theta) \left( 2\beta^{-1}  \frac{\partial^2}{\partial \theta \, \partial \theta^*}
 +  \overrightarrow{\mbox{sig}}_{\theta} \, \frac{\partial}{\partial t}  \right)
 \;, 
\end{eqnarray}
where  $\overrightarrow{\mbox{sig}}_\theta $ is a short notation for $2\theta \frac{\partial }{\partial \theta} - 1 $. It is equal to 1 if there is a $\theta$ factor in the right and to -1 otherwise. $\mathbf{D}^{(2)}$ can be written as
\begin{eqnarray}
  \mathbf{D}^{(2)}(\Upsilon', \Upsilon)  =  \gamma(t'-t) \delta({\theta^*}'-\theta^*) \delta(\theta'-\theta)  
  \left( \bar\mathbf{D} \mathbf{D} -\mathbf{D} \bar\mathbf{D} \right) \;,
\end{eqnarray}
with the (covariant\footnote{Covariant in the sense that the derivative of a supersymmetric 
expression is still  supersymmetric.}) derivatives acting on the superspace:
\begin{eqnarray}
 \bar\mathbf{D} \equiv  \frac{\partial}{\partial \theta}\;, \qquad
\mathbf{D}   \equiv \beta^{-1} \frac{\partial}{\partial \theta^*} - \theta \frac{\partial}{\partial t} \;, 
\end{eqnarray}
that obey\footnote{Therefore the $\dot\Psi^2$ term in ${\cal L}[\Psi]$
can be written in terms of covariant derivatives as
$\left(\{\bar\mathbf{D} ,\mathbf{D}\} \Psi\right)^2$.} $\{
\bar\mathbf{D} ,\mathbf{D}\} = - \frac{\partial}{\partial t}$ and
$\{\mathbf{D}, \mathbf{D} \} = \{\bar\mathbf{D}, \bar\mathbf{D} \}
=0$.
In the white noise limit the dissipative part of the action simplifies to
\begin{eqnarray}
  S_\mathrm{susy}^\mathrm{diss}[\Psi] &=& \frac{1}{2} \int \Ud{\Upsilon} M(\Psi(\Upsilon))\, \mathbf{D}^{(2)}(\Upsilon) \, M(\Psi(\Upsilon))\;,
\end{eqnarray}
with the `dissipative' differential operator
\begin{eqnarray}
 \mathbf{D}^{(2)}(\Upsilon) &\equiv&
\gamma_0 \left( 2\beta^{-1}  \frac{\partial^2}{\partial \theta \, \partial \theta^*}
 +  \overrightarrow{\mbox{sig}}_\theta \, \frac{\partial}{\partial t}  \right)
= \gamma_0 \left( \bar\mathbf{D} \mathbf{D} -\mathbf{D} \bar\mathbf{D} \right) \;.
\end{eqnarray}
This formulation is only suitable in situations in which the applied forces are conservative. The Jacobian term $S^{\cal J}$ contributes to both the
deterministic ($S_\mathrm{susy}^\mathrm{det}$) and the dissipative part
($S_\mathrm{susy}^\mathrm{diss}$) of the action.

\subsubsection{Symmetries.}

In terms of the superfield, the transformation ${\cal T}_{\cal J}(\alpha)$ defined in 
eq.~(\ref{eq:TransfoGhosts}) acts as
\begin{eqnarray}
{\cal T}_{\cal J}(\alpha) \equiv \Psi(t,\theta,\theta^*)  \mapsto \Psi(t, \alpha^{-1} \theta, \alpha\theta^*)
\quad \forall\, \alpha \in \mathbb{C}^*\;, \label{eq:TransGhosts}
\end{eqnarray}
and leaves the action $S[\Psi]$, see eq.~(\ref{eq:SPsi}), invariant.
The transformation ${\cal T}_\mathrm{eq}$ given in eq.~(\ref{eq:TransfoFDT}) acts as
\begin{eqnarray}\label{eq:SUSY_Teq}
{\cal T}_\mathrm{eq} \equiv \Psi(t,\theta,\theta^*)  \mapsto \Psi(-t-\beta\theta^*\theta, -\theta^*, \theta)\;,
\end{eqnarray}
and leaves the action $S[\Psi]$, see eq.~(\ref{eq:SPsi}), invariant. 

The action $S[\Psi]$ given in (\ref{eq:SPsi}) has an additional supersymmetry generated by
\begin{eqnarray}
  \mathbf{Q} \equiv  \frac{\partial}{\partial \theta^*}\;,\qquad
  \bar\mathbf{Q}   \equiv \beta^{-1} \frac{\partial}{\partial \theta} + \theta^* \frac{\partial}{\partial t} \;, 
\end{eqnarray}
that obey $\{ \bar\mathbf{Q} ,\mathbf{Q}\} = \frac{\partial}{\partial
  t}$ and $\{\mathbf{Q} ,\mathbf{Q}\} = \{ \bar\mathbf{Q}
,\bar\mathbf{Q}\} = \{\mathbf{D} ,\mathbf{Q}\} = \{\mathbf{D}
,\bar\mathbf{Q}\} = \{\bar\mathbf{D} ,\mathbf{Q}\} = \{\bar\mathbf{D}
,\bar\mathbf{Q}\} = 0$. Both operators $\mathbf{Q}$ and $\bar\mathbf{Q}$ are thus nilpotent and $\{ \bar\mathbf{Q} ,\mathbf{Q}\}$ is the generator of the Lie sub-group. They act on the superfield as
\begin{eqnarray}
\rme^{\epsilon^*\mathbf{Q}}  \Psi = \Psi +\epsilon^*\mathbf{Q}\Psi\;, \qquad 
\rme^{\epsilon\bar\mathbf{Q}} \Psi = \Psi + \epsilon\bar\mathbf{Q}\Psi\;,
\end{eqnarray}
where $\epsilon$ and $\epsilon^*$ are two extra independent\footnote{$\epsilon$ 
and $\epsilon^*$ are independent of the coordinates $\theta$ and $\theta^*$.} 
Grassmann constants and
\begin{eqnarray}
\mathbf{Q}\Psi &=& c + \theta \left(\rmi\hat\psi + c^*c \frac{M''(\psi)}{M'(\psi)} \right)\;, \\
\bar\mathbf{Q}\Psi &=& -\beta^{-1} c^* - \theta^* \left(\beta^{-1}\rmi\hat\psi - \partial_t \psi + \beta^{-1} c^*c \frac{M''(\psi)}{M'(\psi)} \right) - \theta^*\theta \,\partial_t c^*\;.
\end{eqnarray}
Expressed in terms of superfield transformations, $S[\Psi]$ is invariant under both
\begin{eqnarray}
 \Psi(t,\theta,\theta^*) \mapsto \Psi(t ,\theta ,\theta^* + \epsilon^*) \label{eq:TransBRS}
\end{eqnarray}
and
\begin{eqnarray}
\Psi(t,\theta,\theta^*) \mapsto \Psi(t + \epsilon\theta^*,\theta + \beta^{-1} \epsilon ,\theta^*)\;. \label{eq:TransFDT}
\end{eqnarray}
 
Here again, the invariance of the action is achieved independently by the deterministic 
($S^\mathrm{det}$) and the dissipative ($S^\mathrm{diss}$) contributions.
We would like to stress the fact that the presence of the boundary term accounting for the initial 
equilibrium measure of the field $\psi$ as well as the boundary conditions for the fields  $\rmi\hat
\psi$, $c$ and $c^*$ are necessary to obtain a full invariance of the action.

\subsubsection{BRS symmetry.}\label{sec:BRS} The symmetry generated by $\mathbf{Q}$ is the BRS symmetry that generically arises when a system has dynamical constraints (here we impose the system to obey the Langevin equation of motion). Applying the corresponding superfield transformation in $\langle \Psi(t,\theta,\theta^*) \rangle_{S}$ gives
\begin{eqnarray}
 \langle \Psi(t,\theta,\theta^*) \rangle_{S} = \langle \Psi(t,\theta,\theta^*) + \epsilon^* \mathbf{Q} \Psi(t,\theta,\theta^*) \rangle_{S}\;,
\end{eqnarray}
and therefore $\langle \mathbf{Q} \Psi(t,\theta,\theta^*) \rangle_{S} =0$. This leads to
\begin{eqnarray}\label{eq:Ward1BRS}
 \langle c_t \rangle_{S} 
=0\;,
\qquad
\langle \rmi\hat\psi_t + c^*_t c_t \frac{M''(\psi_t)}{M'(\psi_t)} \rangle_{S} 
= 0\;.
\end{eqnarray}
Applying the transformation inside the two-point correlator $\langle \Psi(t,\theta,\theta^*) \Psi(t',
\theta',{\theta^*}') \rangle_{S}$, we find $\langle  \mathbf{Q}\Psi(t,\theta,\theta^*) \, \Psi(t',\theta',
{\theta^*}')   \rangle_{S} + (t,\theta,{\theta^*})  \leftrightarrow (t',\theta',{\theta^*}')   = 0$. 
This leads to identify the  two-time fermionic correlator as being the (bosonic) linear response:
\begin{eqnarray}
 R(t,t') \equiv \langle \psi_t \left[ \rmi\hat\psi_{t'} + c^*_{t'} c_{t'} \frac{M'(\psi_{t'})}{M''(\psi_{t'})} \right] 
 \rangle_S = \langle c^*_{t'} c_t  \rangle_S\;. \label{eq:BRSRCC}
\end{eqnarray}
Corroborating the discussion in Sect.~\ref{sec:Jaco},  this tells us that $\langle c^*_t 
c_{t'} \rangle_{S}$  (and more generally the fermionic Green function $\langle c^*_t c_{t'} \rangle_
{S^{\cal J}}$) vanishes for $t>t'$ and also for $t=t'$ provided that the Markov limit is not taken. 
Using this result, the second relation in~(\ref{eq:Ward1BRS}) now yields $\langle \rmi\hat\psi_t
\rangle_{S}=0$.

\subsubsection{FDT.}
The use of the symmetry generated by $\bar\mathbf{Q}$ on $\langle \Psi(t,\theta,\theta^*) 
\rangle_{S}$ gives, 
\begin{eqnarray}
 \langle c^*_t \rangle_{S} 
=0\;,
\qquad
 \langle  \rmi\hat\psi_t -  \beta \partial_t   \psi_t \rangle_{S}
= 0\;.
\end{eqnarray}
By use of $\langle\rmi\hat\psi_t\rangle_S = 0$ (which was a consequence of the BRS symmetry), 
the second relation becomes $\partial_t \langle \psi_t \rangle_S = 0$. This expresses the 
stationarity and can be easily generalized to more complicated one-time observables, $A(\psi)$, 
by use of the supersymmetry in $\langle A(\Psi) \rangle_S$.

The use of the symmetry generated by $\bar\mathbf{Q}$ on a two-point correlator of the superfield 
reads
\begin{eqnarray*}
\hspace{-1.0cm} \langle \Psi(t,\theta, \theta^*) \Psi(t',\theta', {\theta^*}') \rangle_{S} = 
\langle \Psi(t+\epsilon\theta^*,\theta+\beta\epsilon, \theta^*) \Psi(t'+\epsilon{\theta^*}',\theta'+\beta
\epsilon, {\theta^*}') \rangle_{S}\;,
\end{eqnarray*}
implying, amongst other relations,
\begin{eqnarray}\label{eq:preFDT}
  \langle \psi_t  \left[    \rmi\hat\psi_{t'} -\beta \partial_{t'} \psi_{t'}  + c^*_t c_t \frac{M''(\psi_t)}
  {M'(\psi_t)} \right]  - c^*_{t} c_{t'} \rangle_{S}  = 0\;.
\end{eqnarray}
As discussed in Sect.~\ref{sec:BRS}, $\langle c^*_{t} c_{t'} \rangle_{S^{\cal J}}$ vanishes for 
$t\geq t'$. Therefore, the term in $c^*_t c_t$ disappears from eq.~(\ref{eq:preFDT}) and 
the FDT is obtained by multiplying both sides of the equation by $\Theta(t-t')$
\begin{eqnarray}
R(t,t') &=& \beta \partial_{t'} C(t,t') \Theta(t-t')\;.
\end{eqnarray}

\subsection{Link between ${\cal T}_\mathrm{eq}$ and the supersymmetries}

It is interesting to remark that both supersymmetries (the one generated by  $\mathbf{Q}$  and the 
one generated by $\mathbf{\bar Q}$) are needed to derive equilibrium relations such as 
stationarity or the FDT. All the Ward-Takahashi identities generated by the combined use of these 
supersymmetries can be generated by ${\mathcal T}_\mathrm{eq}$ but the inverse is not true. The 
supersymmetries do not yield relations in which a time-reversal appears explicitly 
such as the Onsager reciprocal relations.

It is clear from its expression in terms of the superfield, eq.~(\ref{eq:SUSY_Teq}),  that the 
equilibrium transformation ${\mathcal T}_\mathrm{eq}$ cannot be written using the  
generator of a continuous supersymmetry. However, the transformation 
${\cal T}_\mathrm{eq}$ can be formally 
written in terms of the supersymmetry generators as
\begin{eqnarray}
 {\cal T}_\mathrm{eq} \equiv \Psi \mapsto \Pi \, \Xi \, \rme^{\mathbf{\tilde Q}}  \, \Psi\;,
\end{eqnarray}
where $\Pi$ is the time-reversal operator ($t\mapsto-t$), $\Xi$ exchanges the extra Grassmann coordinates ($\theta \mapsto - \theta^*$ and $\theta^* \mapsto \theta$) and the generator $\mathbf{\tilde Q}$ is defined in terms of $\mathbf{Q}$ and $\bar\mathbf{Q}$ as
\begin{eqnarray}\label{eq:deftildeQ}
 \mathbf{\tilde Q} \equiv -\beta \theta^* \theta \, \{ \bar\mathbf{Q} ,\mathbf{Q}\} = -\beta \theta^* \theta \frac{\partial }{\partial t}  \;.
\end{eqnarray}

\section{Out of equilibrium}
\label{sec:outofequilibrium}

We now turn to more generic situations in which the system does no
longer evolve in equilibrium.  This means that it can now be prepared
with an arbitrary distribution and it can evolve with time-dependent
and non-conservative forces $\mathrm{f}$.

We first show that the way in which the symmetry ${\cal T}_\mathrm{eq}$
is broken gives a number of so-called transient\footnote{As opposed to \textit{steady-state} fluctuation relations the validity of which is only asymptotic, in the limit of long averaging times.} fluctuations
relations~\cite{Evans}-\cite{FT-reviews,Mallick}.  Although fluctuation
theorems in cases with additive colored noise were studied in several
publications~\cite{Zamponi-etal}-\cite{Ohta}, we are not aware of
similar studies in cases with multiplicative noise.

We then exhibit another symmetry of the MSRJD generating functional,
valid in and out of equilibrium. This new symmetry implies out of
equilibrium relations between correlations and
responses and generalizes the formul\ae \ in~\cite{Cukupa}-\cite{CorberiLipp}
obtained for additive white noise.  Finally, we come back to
the equilibrium case to combine the two symmetries and deduce other
equilibrium relations.

\subsection{Non-equilibrium fluctuation relations}

\subsubsection{Work fluctuation theorems.}
Let us assume that the system is initially prepared in thermal
equilibrium with respect to the potential
$V(\psi,\lambda_{-T})$\footnote{This is in fact a restriction on the
  initial velocities, $\dot\psi_{-T}$, that are to be taken from the
  Boltzmann distribution with temperature $\beta^{-1}$, independently
  of the positions $\psi_{-T}$. The distribution of the latter can be tailored at
  will through the $\lambda$ dependence of $V$.}.  The expression for
the deterministic part of the MSRJD action functional [see
eq.~(\ref{eq:Sdet})] is
\begin{eqnarray}
 S^\mathrm{det}[\psi, \hat\psi; \lambda, \mathrm{f}] &=&
 -\beta {\cal H}([\psi_{-T}], \lambda_{-T})  - \ln \mathcal{Z}(\lambda_{-T})  \nonumber \\
& & -\int_u  \rmi\hat \psi_u \left[ m\ddot \psi_u + 
 V'(\psi_u,\lambda_u)-\mathrm{f}_u[\psi]  \right] 
 \;,
 \end{eqnarray}
where ${\cal H}([\psi_t], \lambda_t) \equiv  \frac{1}{2} m \dot \psi^2_t  + V(\psi_{t},\lambda_{t})$. 
The external work done on the system along a given trajectory between times $-T$ and
$T$ is the sum of the work induced by the non-conservative forces and the one performed through the external protocol $\lambda$:
\begin{eqnarray}
 W[\psi ; \lambda,\mathrm{f}] \equiv  \int_u \dot \psi_u \ \mathrm{f}_u[\psi] + \int_u  \partial_u \lambda_u \, \partial_\lambda V(\psi_u,\lambda_u) \;. \label{eq:work}
\end{eqnarray}
The equilibrium transformation $\mathcal{T}_\mathrm{eq}$ does not leave $S^\mathrm{det}$ invariant but yields
\begin{eqnarray}
 S^\mathrm{det}[\psi, \hat\psi;  \lambda,\mathrm{f}] \mapsto S^\mathrm{det}[\psi,\hat\psi;  \bar\lambda, \mathrm{f}_\mathrm{r}] + \beta \Delta {\cal F}_\mathrm{r} - \beta
W[\psi;\bar\lambda, \mathrm{f}_\mathrm{r}]\;,
\end{eqnarray}
or equivalently
\begin{eqnarray}
 S^\mathrm{det}[\psi, \hat\psi;  \lambda,\mathrm{f}] + \beta \Delta {\cal F} - \beta
W[\psi;\lambda, \mathrm{f}] \mapsto S^\mathrm{det}[\psi,\hat\psi;  \bar\lambda, \mathrm{f}_\mathrm{r}] \;.
\end{eqnarray}
$S^\mathrm{det}[\psi,\hat\psi; \bar\lambda, \mathrm{f}_\mathrm{r}]$ corresponds to the MSRJD action of the system that is prepared (in equilibrium) and
evolves under the time-reversed protocol $\bar\lambda(u) \equiv
\lambda(-u)$ and external forces $\mathrm{f}_\mathrm{r}([\psi],u) \equiv
\mathrm{f}([\bar \psi],-u)$. $\Delta {\cal F}_\mathrm{r}$ is the change in free
energy associated to this time-reversed protocol: $\beta \Delta {\cal F}_\mathrm{r} = - \ln {\cal Z}(\bar\lambda(T)) + \ln {\cal
  Z}(\bar\lambda(-T)) = -\beta \Delta {\cal F}$ between the initial and the final `virtual'
equilibrium states.  The dissipative part of the action, $S^{\rm
  diss}$, is still invariant under ${\cal T}_\mathrm{eq}$. This means
that, contrary to the external forces $F$, the interaction with the
bath is time-reversal invariant: the friction is still dissipative
after the transformation. This immediately yields
\begin{eqnarray}
 \rme^{\beta \Delta {\cal F}} \langle A[\psi,\hat\psi] \rme^{-\beta W[\psi;\lambda,\mathrm{f}]} \rangle_{S[\lambda, \mathrm{f}]} = 
  \langle A[\mathcal{T}_\mathrm{eq}\psi,\mathcal{T}_\mathrm{eq}\hat\psi]  \rangle_{S[\bar\lambda,\mathrm{f}_\mathrm{r}]}\;
\end{eqnarray}
for any functional $A$ of $\psi$ and $\hat\psi$.
In particular, for a local functional of the field,  $A[\psi(t)]$, it leads to 
the relation~\cite{Crooks2}
\begin{eqnarray}
 \rme^{\beta \Delta {\cal F}}  \langle A[\psi(t)] \rme^{-\beta W[\psi;\lambda ,\mathrm{f}]} \rangle_{S[\lambda, \mathrm{f}]}
= 
\langle A_\mathrm{r}[\psi(-t)] 
\rangle_{S[\bar\lambda,\mathrm{f}_\mathrm{r}]}\;,
\end{eqnarray}
or also
\begin{eqnarray}
 \rme^{\beta \Delta {\cal F}}  \langle A[\psi(t)]  B[\psi(t')]  \rme^{-\beta W[\psi; \lambda ,\mathrm{f}]} \rangle_{S[\lambda, \mathrm{f}]} \nonumber \\
\qquad\qquad= 
 \langle A_\mathrm{r}[\psi(-t)] B_\mathrm{r}[\psi(-t')] 
\rangle_{S[\bar\lambda,\mathrm{f}_\mathrm{r}]}.
\end{eqnarray}
Setting $A[\psi,\hat\psi]=1$, we obtain the Jarzynski equality~\cite{Jarzynski}
\begin{eqnarray}
  \rme^{\beta \Delta {\cal F}}  
  \langle  \rme^{-\beta W[\psi;\lambda,\mathrm{f}]} \rangle_{S[\lambda, \mathrm{f}]} = 1\;.
\end{eqnarray}
Setting $A[\psi,\hat\psi] = \delta(W - W[\psi;\lambda,\mathrm{f}])$
we deduce the Crooks fluctuation theorem~\cite{Crooks1,Kurchan}
\begin{eqnarray}
 P(W) = P_\mathrm{r}(-W) \ \rme^{\beta(W-\Delta {\cal F})}\;,
\end{eqnarray}
where $P(W)$ is the probability for the external work done between
$-T$ and $T$ to be $W$ given the protocol $\lambda(t)$ and the
non-conservative force $\mathrm{f}([\psi],t)$. $P_\mathrm{r}(W)$ is the same
probability, given the time-reversed protocol $\bar\lambda$ and
time-reversed force $\mathrm{f}_\mathrm{r}$. The previous Jarzynski equality is the integral version of this theorem.

\subsubsection{Fluctuation theorem.}
Let us now relax the condition that the system is prepared in thermal equilibrium and allow for any initial distribution $P_{\rmi}$.
We recall the corresponding deterministic part of the MSRJD action functional given in Sect.~\ref{sec:TheGF}, eq.~(\ref{eq:Sdet})
\begin{eqnarray} 
 S^\mathrm{det}[\psi, \hat\psi] 
 &\equiv& 
\ln P_\mathrm{i}\left(\psi(-T),\dot\psi(-T)\right) \nonumber
\\
 & & - \int \Ud{u} \rmi\hat \psi(u) \left[ m\ddot \psi(u) + 
 V'(\psi(u),\lambda(u)) - \mathrm{f}([\psi],u) \right].
\end{eqnarray}
The transformation ${\cal T}_\mathrm{eq}$ does not leave $S^\mathrm{det}$ invariant but one has
\begin{eqnarray}
 S^\mathrm{det}[\psi, \hat\psi;  \lambda,\mathrm{f}] - {\cal S} \mapsto && S^\mathrm{det}[\psi,\hat\psi;  \bar\lambda, \mathrm{f}_\mathrm{r}]\;,
\end{eqnarray}
with the stochastic entropy  ${\cal S} \equiv -\left[ \ln P_\rmi(\psi(T),-\dot\psi(T)) - \ln P_\rmi(\psi(-T),\dot\psi(-T)) \right] - \beta {\cal Q}$.
The first term is the Shannon entropy whereas the second term is the exchange entropy defined through the heat transfer ${\cal Q}
\equiv \Delta {\cal H} - W[\psi;\lambda,\mathrm{f}]$. $\Delta {\cal H} \equiv {\cal H}([\psi(T)],\lambda(T)) - {\cal H}([\psi(-T)],\lambda(-T))$ is the change of internal energy.
The dissipative part of the action, $S^\mathrm{diss}$, is still invariant under ${\cal T}_\mathrm{eq}$. This immediately yields
\begin{eqnarray}
 \langle A[\psi,\hat\psi]\rme^{-{\cal S}}  \rangle_{S[\lambda, \mathrm{f}]} = 
\langle A[\mathcal{T}_\mathrm{eq}\psi,\mathcal{T}_\mathrm{eq}\hat\psi]  \rangle_{S[\bar\lambda,\mathrm{f}_\mathrm{r}]}\;
\end{eqnarray}
for any functional $A$ of $\psi$ and $\hat\psi$.
Setting $A[\psi,\hat\psi]=1$, we obtain the integral fluctuation theorem (sometimes referred as the 
Kawasaki identity)
\begin{eqnarray}
  1 =  \langle  \rme^{-{\cal S}} \rangle_{S[\psi,\hat\psi;
\lambda, \mathrm{f}]}\;,
\end{eqnarray}
which using the Jensen inequality gives $\langle {\cal S} \rangle_{S[\psi,\hat\psi;
\lambda, \mathrm{f}]} \geq 0$, expressing the second law of thermodynamics.
Setting $A[\psi,\hat\psi] = \delta(\zeta - {\cal S})$
we derive the fluctuation theorem~\cite{Gallavotti,Kurchan}
\begin{eqnarray}
 P(\zeta) = P_\mathrm{r}(-\zeta) \ \rme^{\zeta}\;,
\end{eqnarray}
where $P(\zeta)$ is the probability 
for the entropy created between
$-T$ and $T$ to be $\zeta$ given the protocol $\lambda(t)$ and the
non-conservative force $\mathrm{f}([\psi],t)$. $P_\mathrm{r}(\zeta)$ is the same
probability, given the time-reversed protocol $\bar\lambda$ and
time-reversed force $\mathrm{f}_\mathrm{r}$.

Similar results can be obtained for isolated systems by switching off the 
interaction with the bath, \textit{i.e.} by taking $\gamma=0$. It is also straightforward
to obtain extended relations when the bath is taken to be out of equilibrium, for
example by using $\Gamma(t-t')\neq\gamma(t-t')+\gamma(t'-t)$, and the 
contribution of the change in the dissipative action is taken into account. 
This kind of fluctuation relation may be specially important in quantum systems.

\subsection{Generic relations between correlations and linear responses}

A number of generic relations between linear responses and the
averages of other observables have been derived for different types of
stochastic dynamics: Langevin with additive white noise~\cite{Cukupa},
Ising variables with Glauber updates~\cite{Chatelain}, or the
heat-bath algorithm~\cite{Ricci-Tersenghi,Corberi,Diezemann,Crri}, and
even molecular dynamics of hard spheres or Lennard-Jones particle
systems~\cite{Berthier}. Especially interesting are those in which the
relation is established with functions of correlations computed with
the unperturbed dynamics~\cite{Cukupa,Corberi} as explained in
\cite{CorberiLipp}. The main aim of the studies
in~\cite{Chatelain}-\cite{CorberiLipp} was to give the most efficient
computational method to obtain the linear response in the theoretical
limit of no applied field. Another set of recent articles discusses 
very similar  relations with the
goal of giving a thermodynamic interpretation to the various
terms contributing to the linear response~\cite{Baesi}-\cite{Villamaina}.

In the concrete case of Langevin processes this kind of relations can be very
simply derived by multiplying the equation by the field or the noise
and averaging over the noise in the way done in~\cite{Cukupa}.
We derive here the same relations within the 
MSRJD formalism, using a symmetry property that is more likely to
admit an extension to systems with quantum fluctuations.

\subsubsection{A symmetry of the MSRJD generating functional valid also out of equilibrium.}
We consider the most generic out of equilibrium situation. We allow for any initial preparation ($P_\rmi$) and any evolution of the system ($F$).
$\int \uD{[\psi,\hat\psi]}  \rme^{S[\psi,\hat\psi]} \nonumber$
is invariant under the involutary field transformation $\mathcal{T}_\mathrm{eom}$, given by
\begin{equation}
\mathcal{T}_\mathrm{eom} \equiv
\left\{
\begin{array}{rcl}
 \psi_u &\mapsto& \psi_u\;, \\
\rmi\hat\psi_u
&\mapsto& \displaystyle
-\rmi{\hat\psi}_u + 
\frac{2\beta}{M'(\psi_u)} \int_v \Gamma^{-1}_{u-v \,}\, \frac{\mbox{\small\sc{Eq}}_v[\psi]}{ M'(\psi_v) }
\; , 
\end{array}
\right. \label{eq:TransfoGen}
\end{equation}
The meaning of the subscript referring to `equation of motion' will become clear in the following.
For additive noise [$M'(\psi) = 1$] the transformation becomes
\begin{eqnarray*}
\rmi\hat \psi_u  \mapsto
-\rmi{\hat\psi}_u + 
2\beta 
 \int_v \Gamma^{-1}_{u-v} 
\left[ m\ddot \psi_v -  F_v[\psi] + 
\int_w \gamma_{v-w} \dot \psi_w \right] \;,
\end{eqnarray*}
and in the white noise limit it simplifies to
\begin{eqnarray}
 \rmi\hat \psi_u 
&\mapsto& 
-\rmi{\hat\psi}_u + 
\beta \gamma_0^{-1} \left[ m\ddot \psi_u -  F_u[\psi] + \gamma_0 \dot \psi_u \right] \;.
\end{eqnarray}
The proof of invariance is similar to the one developed in Sect.~\ref{sec:FDTSym}
when dealing  with the equilibrium symmetry.  The Jacobian of this transformation is unity
since its associated matrix is block triangular with ones on the
diagonal. The integration domain of $\psi$ is unchanged while the one
of $\hat\psi$ can be chosen to be the real axis by a simple complex
analysis argument. In the following lines we show that the action
$S$ evaluated in the transformed fields remains identical to
the action evaluated in the original fields. We give the proof using an additive 
noise but the generalization to a multiplicative noise is straightforward.
We start from the
expression~(\ref{eq:Action_Fancy}) and evaluate
\begin{eqnarray}
S[\mathcal{T}_\mathrm{eom} \psi, \mathcal{T}_\mathrm{eom}\hat\psi] 
=
\ln P_\mathrm{i}(\psi_{-T},\dot\psi_{-T})
+ 
\int_{u}  
\left[ \rmi\hat\psi_u -  2\beta \int_{v} \Gamma^{-1}_{u-v} \mbox{\small\sc{Eq}}_{v}[\psi] \right]
\nonumber \\
\qquad\qquad\qquad \times
\left[ \mbox{\small\sc{Eq}}_u[\psi] - \frac{1}{2} \int_{w} \beta^{-1}\Gamma_{u-w} \left(-\rmi\hat\psi_{w} + 
2\beta \int_{z} \Gamma^{-1}_{w-z} \mbox{\small\sc{Eq}}_{z}[\psi]\right)\right]
\nonumber\\
\quad=
\ln P_\mathrm{i}(\psi_{-T},\dot\psi_{-T})
+ 
\int_{u}  
\left[ \rmi\hat\psi_u -  2\beta \int_{v} \Gamma^{-1}_{u-v} \mbox{\small\sc{Eq}}_{v}[\psi] \right]
\nonumber
\left[ \frac{1}{2} \int_{w} \beta^{-1}\Gamma_{u-w} \ \rmi\hat\psi_{w} \right]
\nonumber\\
\quad = S[\psi, \hat\psi] 
\; . 
\end{eqnarray}
Contrary to the equilibrium transformation $\mathcal{T}_\mathrm{eq}$, it does not include a time-reversal and is not defined in the Newtonian limit ($\gamma=0$).

\subsubsection{Supersymmetric version.}
In Sect.~\ref{sec:susy}, we encoded the fields $\psi$, $\rmi\hat\psi$, $c$ 
and $c^*$ in a unique superfield $\Psi$. In this fashion, the transformation ${\cal T}_\mathrm{eom}$ given in 
eq.~(\ref{eq:TransfoGen}) acts as
\begin{eqnarray}
\Psi(t,\theta,\theta^*)  \mapsto \Psi\left(t + \theta^*\theta \,\frac{2 \beta \int_u  \Gamma^{-1}_{t-u} 
M'(\Psi(u,\theta,\theta^*)) \mbox{\small\sc{Eq}}_u[\Psi]}{\partial_t M(\Psi(t, \theta, \theta^*))}, \theta, 
\theta^*\right)\;,
\end{eqnarray}
and leaves the equilibrium action $S[\Psi]$, see eq.~(\ref{eq:SPsi}), invariant.

\subsubsection{Out of equilibrium relations.}\label{sec:o-e-relations}
We first derive some relations in the additive case [$M'(\psi) = 1$] and then we generalize the results to the case of multiplicative noise. 
\paragraph{Additive noise.}
Using ${\cal T}_\mathrm{eom}$ in the expression~(\ref{eq:response-psihat}) of the
self response $R(t,t')$ we find
\begin{eqnarray*}
\langle \psi_t \rmi\hat\psi_{t'} \rangle_{S} 
=
 \langle \mathcal{T}_\mathrm{eom} \psi_t \, \mathcal{T}_\mathrm{eom}
\rmi\hat\psi_{t'}  \rangle_{S} 
=
- \langle \psi_t \rmi\hat\psi_{t'} \rangle_{S}  
+
2 \beta \int_v \Gamma^{-1}_{t'-v} \langle \psi_t  \mbox{\small\sc{Eq}}_{v}[\psi]   \rangle_{S} 
\; , 
\end{eqnarray*}
giving an explicit formula for computing the linear response without perturbing field:
\begin{eqnarray}
&& 
R(t,t') 
= \beta \int \Ud{v} \Gamma^{-1}(t'-v) 
\nonumber\\
&& \quad \times
\left[ m \partial_{v}^2 C(t,v) + \int \Ud{u} \gamma(v-u) \partial_{u} 
C(t,u) - \langle \psi(t) F([\psi],v) \rangle
\right]
\; .
 \label{eq:Schwinger-Dyson0} 
\end{eqnarray}
Once multiplied by $\Gamma_{t''-t'}$ and integrated over $t'$ yields 
\begin{eqnarray}
 m \partial_{t'}^2 C(t,t')
&+& \int \ud{u} \gamma(t'-u) \partial_{u} 
C(t,u) \nonumber \\
&-& \langle \psi(t) F([\psi],t') \rangle
= \beta^{-1} \int \ud{u} \Gamma(t'-u) 
R(t,u) \;, \label{eq:Schwinger-Dyson}
\end{eqnarray}
with no assumption on the initial $P_\mathrm{i}(\psi_{-T},\dot\psi_{-T})$.

Using now 
${\cal T}_\mathrm{eom}$ in $\langle  \mbox{\small\sc{Eq}}_{t}[\psi] \rmi\hat\psi_{t'} \rangle_S$, we get
\begin{eqnarray*}
 \langle  \mbox{\small\sc{Eq}}_{t}[\psi] \rmi\hat\psi_{t'} \rangle_S &=& \langle  \mbox{\small\sc{Eq}}_{t}[\mathcal{T}_\mathrm{eom} \psi] \, \mathcal{T}_\mathrm{eom} \rmi\hat\psi_{t'} \rangle_S \\
&=& -\langle  \mbox{\small\sc{Eq}}_{t}[\psi] \rmi\hat\psi_{t'} \rangle_S + 2\beta \int_u \Gamma^{-1}_{t'-u} \langle \mbox{\small\sc{Eq}}_{t}[\psi] \mbox{\small\sc{Eq}}_{u}[\psi] \rangle_S\;.
\end{eqnarray*}
Since $\langle \mbox{\small\sc{Eq}}_{t}[\psi] \mbox{\small\sc{Eq}}_{u}[\psi]\rangle_S = \beta^{-1} \Gamma_{t-u}$, this simplifies in
\begin{eqnarray*}
\langle  \mbox{\small\sc{Eq}}_{t}[\psi] \rmi\hat\psi_{t'} \rangle_S = \delta_{t-t'}\;,\nonumber
\end{eqnarray*}
that yields 
\begin{eqnarray}
m \partial^2_t R(t,t') &+& \int \ud{v} \gamma(t-v) \partial_v R(v,t') - 
 \langle \rmi\hat \psi(t') F([\psi],t) \rangle_S 
=\delta(t-t')
\label{eq:Schwinger-Dyson2}
\end{eqnarray}
with no assumption on the initial $P_\mathrm{i}$. One can trade the last term in the {\sc lhs} for 
$\beta \int_u \Gamma^{-1}_{t'-u} \langle \xi(u) F_{t}[\psi] \rangle_{\xi}$ by use of Novikov's theorem.

Integrating both eqs.~(\ref{eq:Schwinger-Dyson}) and
(\ref{eq:Schwinger-Dyson2}) around $t=t'$ we find the equal-time
conditions
\begin{eqnarray}
\hspace{-0.5cm}  m \left. \partial_{t'}C(t,t')\right|_{t' = t} = 0, \quad m\left.\partial_{t}R(t,t')\right|_{t'\to t^-} = 1, \quad m\left.\partial_{t}R(t,t')\right|_{t' \to t^+} = 0\;. \label{eq:eqtcond}
\end{eqnarray}
The last two relations  imply that the first derivative of the
response function is discontinuous at equal times\footnote{It is clear
  from the expressions given in~(\ref{eq:eqtcond}) that the overdamped
  $m\to0$ limit allows for a sudden discontinuity of the response
  function as well as a finite slope of the correlation function at
  equal times.}.

The use of this symmetry is an easy way to get a generalization of
eq.~(\ref{eq:Schwinger-Dyson0}) for a generic response
$R_{AB}$. Indeed, applying this transformation to
expression~(\ref{eq:RabCl}) of the linear response we obtain
\begin{eqnarray}
 R_{AB}(t,t') &=& \beta \int \Ud{u} \Gamma^{-1}(t'-u) \sum_{n=0}^{\infty} \left\{ 
 m \ \partial_u^{n+2} \langle A[\psi(t)] \psi(u) \frac{\partial B[\psi(t')]}{\partial \ \partial_{t'}^{n}\psi(t')} \right. \rangle_{S} \nonumber \\
& &  \qquad - \partial_u^{n}  \langle A[\psi(t)] F([\psi],u) \frac{\partial B[\psi(t')]}{\partial \ \partial_{t'}^{n}\psi(t')} \rangle_{S} \nonumber \\
& & \qquad + \left. \int \Ud{v} \gamma(u-v) \partial_v^{n+1} \langle A[\psi(t)] \psi(v) \frac{\partial B[\psi(t')]}{\partial \ \partial_{t'}^{n} \psi(t')} \rangle_{S}
 \right\}\;.\label{eq:Rab_super}
\end{eqnarray}
This formula gives the linear response as an explicit function of
multiple-time correlators of the field $\psi$. For example, if $B$ is a 
function of the field only (and not of its time-derivatives), just
the $n=0$-term subsists in the above sum:
\begin{eqnarray}
  R_{AB}(t,t') &=& \beta \int \Ud{u} \Gamma^{-1}(t'-u) \left\{ 
    m \ \partial_u^{2} \langle A[\psi(t)] \psi(u) \frac{\partial B[\psi(t')]}{\partial \ \psi(t')} \right. \rangle_{S} \nonumber \\
  & & \qquad  -  \langle A[\psi(t)] F([\psi],u) \frac{\partial B[\psi(t')]}{\partial \ \psi(t')} \rangle_{S} \nonumber \\
  & & \qquad  + \left. \int \Ud{v} \gamma(u-v) \partial_v \langle A[\psi(t)] \psi(v) \frac{\partial B[\psi(t')]}{\partial \ \psi(t')} \rangle_{S}
  \right\}\;.
\end{eqnarray}
As another example if one is interested in the self-response of the
velocity, $A[\psi(t)] = B[\psi(t)] = \partial_t\psi(t)$, one obtains
\begin{eqnarray}
  R_{AB}(t,t') = \beta \int \Ud{u} \Gamma^{-1}(t'-u) \left\{ \begin{array}{l} \\ \end{array} \right. \hspace{-1em}
  & &  m \ \partial_{t} \partial_u^{3} C(t,u) 
  - \partial_{t} \partial_u  \langle  \psi(t) F([\psi],u) \rangle_{S}  \nonumber \\
  & &  + \int \Ud{v} \gamma(u-v) \partial_v^{2} C(t,v)  \hspace{-1em} \left. \begin{array}{l} \\ \end{array} \right\}\;.
\end{eqnarray}

\paragraph{Multiplicative noise.} Similar results can be obtained for multiplicative noise.
Applying the transformation to the correlator $\int_u   \Gamma_{t'-u} \langle \psi_t M'(\psi_{t'}) M'(\psi_{u}) \rmi\hat\psi_{u} \rangle_{S}$ we obtain
\begin{eqnarray}
\langle \psi_t  \mbox{\small\sc{Eq}}_{t'}[\psi]  \rangle_{S} 
= 
\beta^{-1}
 \int_u   \Gamma_{t'-u} \langle \psi_t M'(\psi_{t'}) M'(\psi_{u}) \rmi\hat\psi_{u} \rangle_{S} 
\; , 
\nonumber
\end{eqnarray}
implying
\begin{eqnarray}
 && m \partial_{t'}^2 C(t,t')
 +\int_u \gamma_{t'-u} \langle \psi_t M'(\psi_{t'}) M'(\psi_u) \partial_{u} \psi_u \rangle_S
 \nonumber \\
&& 
\qquad
-
\langle \psi_t F_{t'}[\psi] \rangle_S
= \beta^{-1} \int_u   \Gamma_{t'-u} \langle \psi_t M'(\psi_{t'}) M'(\psi_u) \rmi\hat\psi_u \rangle_{S} 
\;.
\label{eq:Schwinger-DysonM}
\end{eqnarray}
Applying now the transformation to the correlator $\langle  \mbox{\small\sc{Eq}}_{t}[\psi] \rmi\hat\psi_{t'} \rangle_S$, one obtains
\begin{eqnarray}
 \langle  \mbox{\small\sc{Eq}}_{t}[\psi] \rmi\hat\psi_{t'} \rangle_S = \delta_{t-t'} + \beta^{-1} \int_u \Gamma_{t-u}
\langle M'(\psi_t) M'(\psi_u) \rmi\hat\psi_u  \rmi\hat\psi_{t'}  \rangle_S
\end{eqnarray}
yielding
\begin{eqnarray}
 && 
 m \partial_{t}^2 R(t,t')
+ \int_u \gamma_{t-u} \langle  M'(\psi_{t'}) M'(\psi_u) \partial_{u} \psi_u \rmi\hat\psi_{t'} \rangle_S
 \nonumber \\
&&
\qquad 
- \langle  F_{t}[\psi] \rmi\hat\psi_{t'}  \rangle_S
= \delta_{t-t'} \beta^{-1} \int_u   \Gamma_{t-u} \langle \psi_t M'(\psi_{t'}) M'(\psi_u) \rmi\hat\psi_u \rangle_{S} 
\;.
\label{eq:Schwinger-DysonM2}
\end{eqnarray}
One can check from eqs.~(\ref{eq:Schwinger-DysonM}) and (\ref{eq:Schwinger-DysonM2}) that 
the equal-time conditions given in eqs.~(\ref{eq:eqtcond}) are still valid in the multiplicative case.

\subsection{Composition of $\mathcal{T}_\mathrm{eom}$ and $\mathcal{T}_\mathrm{eq}$}
For an equilibrium situation, the MSRJD action functional is fully invariant under the
composition of $\mathcal{T}_\mathrm{eom}$ and $\mathcal{T}_\mathrm{eq}$,
\begin{eqnarray*}
 \mathcal{T}_\mathrm{eq} \circ \mathcal{T}_\mathrm{eom} =
\left\{
 \begin{array}{rcl}
\psi_u  &\mapsto&  \psi_{-u} \;, \\ 
\rmi\hat \psi_u &\mapsto& \displaystyle -\rmi\hat\psi_{-u} - \beta \partial_{u} \psi_{-u} + \frac{2\beta}
{M'(\psi_{-u})} \int_v \Gamma_{u-v}^{-1} \frac{\mbox{\small\sc{Eq}}_{v}[\bar\psi] }{M'(\psi_{-v}) } \;,
\end{array}
\right.
\end{eqnarray*}
that simply reads in the white noise limit
\begin{eqnarray*}
\mathcal{T}_\mathrm{eq} \circ \mathcal{T}_\mathrm{eom} =
\left\{
 \begin{array}{rcl}
\psi_u  &\mapsto&  \psi_{-u} \;, \\ 
\rmi\hat \psi_u  &\mapsto& \displaystyle -\rmi\hat\psi_{-u} +  \frac{\beta}{\gamma_0 M'(\psi_{-u})^2} \left[ m \partial_{u}^2 \psi_{-u} + V'(\psi_{-u}) \right]\;.
\end{array}
\right.
\end{eqnarray*}
For simplicity we only show the implication of this symmetry in this limit and in the additive noise case:
\begin{eqnarray}
 R(t,t')
&=&
- R(-t,-t')
+
 \frac{\beta}{\gamma_0} \left[ m \partial_{t'}^2 C(-t,-t') + \langle \psi(-t) V'(\psi(-t')) \rangle_{S}  \right]
\; .
\nonumber
\end{eqnarray}
Using equilibrium properties, \textit{i.e.} time-translational
invariance of all observables and time-reversal symmetry of two-time
correlation functions of the field $\psi$ (shown in
Sect.~\ref{sec:FDTSym}), and causality of the response, 
\begin{eqnarray}
 R(\tau) = \Theta(\tau) \frac{\beta}{\gamma_0}\left[ m \partial_{\tau}^2 C(\tau) +  
  \langle \psi(t)
V'(\psi(t')) \rangle_{S}\right]\;,
\end{eqnarray}
with $\tau\equiv t-t'$ which is eq.~(\ref{eq:Schwinger-Dyson})
after cancellation of the {\sc lhs} with the last term in the {\sc
  rhs} when FDT between $R$ and $C$ holds [also
eq.~(\ref{eq:Schwinger-Dyson2}) after a similar simplification].  Here
again, one can easily obtain a generalization of this last relation
for a generic response $R_{AB}$ by plugging the transformation into
the expression~(\ref{eq:RabCl}) of the linear response.

\section{Conclusions}
\label{sec:conclusions}

In this paper we recalled the path-integral approach to classical stochastic
dynamics with generic multiplicative colored noise. The action has
three terms: a deterministic (Newtonian dynamics) contribution, a
dissipative part and a Jacobian. We identified
a number of symmetries of  the generating functional when the
sources are set to zero.  The invariance of the action is achieved by
the three terms independently.

One of these symmetries applies only when equilibrium dynamics are
assumed.  Equilibrium dynamics are ensured whenever the system is prepared
with equilibrium initial conditions at temperature $\beta^{-1}$ (a
statistical mixture given by the Gibbs-Boltzmann measure),
evolves with the corresponding time-independent conservative forces, and
is in contact with an equilibrium bath at the same temperature
$\beta^{-1}$. The invariance also holds in the limit in which the contact
with the bath is suppressed, \textit{ i.e.} under deterministic (Newtonian)
dynamics, but the initial condition is still taken from the 
Gibbs-Boltzmann measure. This symmetry
yields all possible model-independent fluctuation-dissipation theorems
as well as stationarity and Onsager reciprocal relations. When the field-transformation is 
applied to driven problems, the symmetry no longer holds, 
but it gives rise to different kinds of fluctuation
theorems.

We identified another more general symmetry that applies to
equilibrium and out of equilibrium set-ups. It holds for any kind
of initial conditions -- they can be any statistical mixture or even
deterministic, and the evolution can be dictated by 
time-dependent and/or non-conservative forces as long as the 
system is coupled to an equilibrium bath. The symmetry implies exact dynamic
equations that couple generic correlations and linear responses. These
equations are model-dependent in the sense that they depend explicitly
on the applied forces. They are the starting point to derive
Schwinger-Dyson-type approximations and close them on two-time
observables.  Although the symmetry is ill-defined in the Newtonian limit,
the dynamic relations it yields can nevertheless be evaluated in the
Newtonian case.

Finally, we gave a supersymmetric expression of the path-integral for
problems with multiplicative colored noise and conservative forces.
We expressed all the previous symmetries in terms of superfield transformations and
we discussed the relationship between supersymmetry and other symmetries.

We intended to present a self-contained presentation of some symmetry
properties of classical deterministic and stochastic dynamics.
We focused on the so-called model A dynamics (with a non-conserved order parameter) for
 a 0-dimensional field. The generalization to a vector field is straightforward. 
Extensions to this work include the study of other dynamics such as the so-called model B
dynamics (with a conserved order parameter). Higher dimensional fields would also allow the study of
forces that do not work, such as Larmor precession around a magnetic vector field~\cite{Bausch}.
This article should serve as an introduction to and motivation for the study of
quantum problems that we shall develop in~\cite{Arbicu}.  Although some  of
these results were known, notably those associated to additive noise
processes, they were scattered and somehow hidden in different
publications. The close relation between all these properties was not 
always fully appreciated either. The multiplicative noise results are, as far as we know,
new.

\vspace{1cm}
\noindent
\underline{Acknowledgments.}
We thank J. Kurchan, F. van Wijland and J. B. Zuber for useful discussions.
This research was financially supported by ANR-BLAN-0346 (FAMOUS)
and it was supported in part by the National Science Foundation
under Grant No. NSF PHY05-51164.

\appendix
\section{\hspace{1.8cm} Conventions and notations}
\label{sec:conventions}

$\Theta$ is the Heaviside step function. When dealing with 
 Markov Langevin equations, the choice of the value of the Heaviside step function 
$\Theta(t)$ at $t=0$ is imposed by the choice of the 
It\^o [$\Theta(0)=0$] or the Stratonovich convention 
[$\Theta(0)=1/2$]. However, away from the Markov case, \textit{i.e.} as long as both inertia and the color of the bath are not neglected simultaneously, the choice of $\Theta(0)$ is unconstrained and the 
physics should not depend on it.
We recall the identities
\begin{equation}
 \int_{-\infty}^{\infty} \frac{\ud{x}}{2\pi} \rme^{ixy}  = \delta(y)
\qquad \qquad 
\mbox{and}
\qquad \qquad 
 \int_{-\infty}^y \Ud{x} \delta(x)  = \Theta(y)
\;,
\end{equation}
where $\delta$ is the Dirac delta function.

\paragraph{Field theory notations.}

Let $\psi$ be a real field. The integration over this field is denoted
$\int \uD{[\psi]}$. If $A$ is a functional of the field, we denote it
$A[\psi]$. If it also depends on one or several external
parameters, such as the time $t$ and a protocol $\lambda$, we denote it
$A([\psi],\lambda,t)$. Whenever $A$ is a local functional of the
field at time $t$ (\textit{i.e.} a function of $\psi(t)$ and its first
time-derivatives), we use the short-hand notation $A[\psi(t)]$. The
time-reversed field constructed from $\psi$ is denoted $\bar\psi$:
$\bar\psi(t)\equiv\psi(-t)$. The time-reversed functional constructed
from $A([\psi],\lambda,t)$ is called $A_\mathrm{r}$: $A_{\rm
  r}([\psi],\lambda,t) \equiv A([\bar\psi],\lambda,-t)$. 
Applied on local observables of $\psi$, it has the effect of changing the sign of all odd time-derivatives in the expression of $A$.

To shorten expressions, we adopt a notation in which the arguments
of the fields appear as subindices, $\psi_t\equiv\psi(t)$, 
$\gamma_{t-t'}\equiv\gamma(t-t')$, and so on and so forth, 
and the integrals over time as
expressed as $\int_t\equiv\int \ud{t}$.

\paragraph{Grassmann numbers.} Let $\theta_1$ and $\theta_2$ be two
anticommuting Grassmann numbers and $\theta_1^*$ and $\theta_2^*$ their respective
Grassmann conjugates. We adopt the following convention for the
complex conjugate of a product of Grassmann numbers: $(\theta_1
\theta_2)^* =\theta_2^* \theta_1^*$.

\section{\hspace{1.8cm} Discrete MSRJD for additive noise}\label{app:MSRJD}

In this Appendix we discuss the MSRJD action for processes with
additive colored noise. 

\subsection{\hspace{1.7cm}Discrete Langevin equation}

The Langevin equation is a stochastic differential equation and 
one can give a rigorous meaning to it by specifying a particular
discretization scheme.

Let us divide the time interval $[-T,T]$ into $N+1$ infinitesimal
slices of width $\epsilon\equiv 2T/(N+1)$. The discretized times are
$t_k = -T + k \epsilon$ with $k=0, ...,N+1$. The discretized version
of $\psi(t)$ is $\psi_k \equiv \psi(t_k)$. The continuum limit is
achieved by sending $N$ to infinity and keeping $(N+1) \epsilon=2T$
constant.  Given some initial conditions $\psi_\mathrm{i}$ and
$\dot\psi_\mathrm{i}$, we set $\psi_1 = \psi_\mathrm{i}$ and $\psi_0 =
\psi_\mathrm{i} - \epsilon \dot\psi_\mathrm{i}$ meaning that the first two
times ($t_0$ and $t_1$) are reserved for the integration over the
initial conditions whereas the $N$ following ones correspond to the
stochastic dynamics given by the discretized Langevin equation:
\begin{eqnarray}
\mbox{\sc{Eq}}_{k-1} &\equiv& 
m \frac{\psi_{k+1} - 2 \psi_{k}  + \psi_{k-1} }{\epsilon^2}  -
 {F_{k}}(\psi_{k}, \psi_{k-1}, ...) 
+  \epsilon \sum_{l=1}^{k}  \gamma_{kl}   
\frac{\psi_{l}-\psi_{l-1}}{\epsilon} 
\nonumber\\
&=& \xi_{k} \;, \label{eq:Langevin_discr}
\end{eqnarray}
defined for $k=1, ..., N$.
The force $F_{k}$ typically depends on the state $\psi_{k}$ but can have a memory kernel (\textit{i.e.} it can depend on previous states $\psi_{k-1}$, $\psi_{k-2}$, etc.). 
The notation $\gamma_{kl}$ stands for
$\gamma_{kl} \equiv \epsilon^{-1} \int_{0^-}^{\epsilon} \Ud{u}
\gamma(t_k - t_l + u)$. The $\xi_k$ are independent
Gaussian random variables with variance $\langle \xi_k \xi_l \rangle =
\beta^{-1} \Gamma_{kl}$ where $\Gamma_{kl} \equiv \gamma_{kl} +
\gamma_{lk} $. Inspecting the equation above, we notice that the value
of $\psi_k$ depends on the realization of the previous noise
realization $\xi_{k-1}$ and there is no need to specify $\xi_0$
and $\xi_{N+1}$. 

In the white noise limit, one has $\gamma_{kl} =
\epsilon^{-1} \gamma_0 \delta_{kl}$, $\langle \xi_k \xi_l \rangle = 2
\gamma_0 \beta^{-1} \epsilon^{-1} \delta_{kl}$ where $\delta$ is the
Kronecker delta, and
\begin{eqnarray}
\mbox{\sc{Eq}}_{k-1} &\equiv& 
m \frac{\psi_{k+1} - 2 \psi_{k}  + \psi_{k-1} }{\epsilon^2}  - 
F_{k}(\psi_{k}, \psi_{k-1}, ...) 
+ \gamma_0  \frac{\psi_{k}-\psi_{k-1}}{\epsilon} 
= \xi_{k} \;.
\nonumber
\end{eqnarray}

\subsection{\hspace{1.7cm}Construction of the MSRJD action}\label{app:MSRJD-Construct}

The probability density $P$ for a complete field history $(\psi_0,
\psi_1, ..., \psi_{N+1})$ is set by the relation
\begin{eqnarray*}
 P(\psi_0, \psi_1, ..., \psi_{N+1}) \udm{\psi_0}\udm{\psi_1} ...\udm{\psi_{N+1}} \nonumber \\
\qquad = 
P_\mathrm{i}(\psi_\mathrm{i},\dot \psi_\mathrm{i}) \udm{\psi_\mathrm{i}} \udm{\dot \psi_\mathrm{i}} \
P_\mathrm{n}(\xi_1, \xi_2, ..., \xi_{N}) \udm{\xi_1}\udm{\xi_2} ...\udm{\xi_{N}} \;. \label{eq:completehist}
\end{eqnarray*}
$P_\mathrm{i}$ is the initial probability distribution of the field. The probability for a given noise history to occur between times $t_1$ and $t_N$ is given by
\begin{eqnarray}\label{eq:noisediscrete}
 P_\mathrm{n}(\xi_1, ..., \xi_N) &=& {\cal M}_N^{-1} \
  \rme^{-\frac{1}{2} \sum_{k,l=1}^{N} {
 \xi_k\, {\beta}\Gamma^{-1}_{kl} \, \xi_l}}
\end{eqnarray}
where $\Gamma^{-1}_{kl}$ is the 
inverse matrix of $\Gamma_{kl}$ (and not the discretized version of the inverse operator of $\Gamma$)
and the
normalization is given by ${\cal M}_N^2 \equiv
\frac{(2\pi)^N}{\mbox{det}_{kl} \left( \beta
    \Gamma^{-1}_{kl} \right) }$ where $\mbox{det} \left( ... \right)$ stands for the matrix determinant. From eq.~(\ref{eq:completehist}),
one derives
\begin{eqnarray} \label{eq:Ppsidiscr} P(\psi_0, \psi_1, ...,
  \psi_{N+1}) = |{\cal J}_N| P_\mathrm{i}(\psi_1, \frac{\psi_1 -
    \psi_0}{\epsilon}) P_\mathrm{n}(\mbox{\small\sc{Eq}}_0, ...,
  \mbox{\small\sc{Eq}}_{N-1}) \;,
\end{eqnarray}
with the Jacobian
\begin{eqnarray*}
  {\cal J}_N \equiv \mbox{det} \left(  \frac{\partial \, (\psi_\mathrm{i},\dot\psi_\mathrm{i}, \xi_1, \ldots, \xi_N )}{\partial \, (\psi_0,\psi_1,\ldots,\psi_{N+1})} \right)
  = \mbox{det} \left( \frac{\partial \, (\psi_\mathrm{i},\dot\psi_\mathrm{i}, \mbox{\small\sc{Eq}}_0, \ldots, \mbox{\small\sc{Eq}}_{N-1} )}{\partial \, (\psi_0,\psi_1,\ldots,\psi_{N+1})} \right) \,, \label{eq:jac_discr_def}
\end{eqnarray*}
that will be discussed in \ref{app:Jacobian}.  The
expression~(\ref{eq:noisediscrete}) for the noise history probability
reads, after a Hubbard-Stratonovich transformation that introduces
the auxiliary variables $\hat\psi_k$ ($k=1, ..., N$),
\begin{eqnarray}
 {\cal N}_N P_\mathrm{n}(\xi_1, ..., \xi_N) =\int \Udm{\hat\psi_1}...\ud{\hat\psi_N} 
 \rme^{ - \epsilon\sum_k \rmi\hat \psi_k \xi_k + \frac{1}{2} {\beta^{-1}} 
 \epsilon^2 \sum_{kl} \rmi\hat\psi_k \Gamma_{kl} \rmi\hat\psi_l} 
\nonumber \\
  \quad = 
  \int \Udm{\hat\psi_0}...\ud{\hat\psi_{N+1}} \delta(\hat\psi_0)\delta(\hat\psi_{N+1})\, 
  \rme^{ - \epsilon\sum_k \rmi\hat\psi_k \mbox{\scriptsize \sc{Eq}}_{k-1} + \frac{1}{2} {\beta^{-1}} \epsilon^2 \sum_{kl} 
  \rmi\hat\psi_k \Gamma_{kl} \rmi\hat\psi_l}  \;, 
\label{eq:noisediscreteHubb}
\end{eqnarray}
with ${\cal N}_N \equiv (2\pi/\epsilon)^N$. In the last step, we replaced
$\xi_k$ by $\mbox{\small\sc{Eq}}_{k-1}$ and we allowed for summations over $k = 0$ and $k=N+1$ as well as integrations over
$\hat\psi_0$ and $\hat\psi_{N+1}$ at the cost of introducing delta
functions. The  Hubbard-Stratonovich transformation allows
for some freedom in the choice of the sign in front of
$\rmi\hat\psi_k$ in the exponent (indeed $P_\mathrm{n}$ is real so
$P_\mathrm{n} = P_\mathrm{n}^*$). Together with eq.~({\ref{eq:Ppsidiscr}})
this gives
\begin{eqnarray}
{\cal N}_N   P(\psi_0, \psi_1, ..., \psi_{N+1})  = \left|{\cal J}_N\right| \int \Udm{\hat\psi_0}...\ud{\hat\psi_{N+1}}  \delta(\hat\psi_0) \delta(\hat\psi_{N+1})  
  \nonumber\\
  \qquad\qquad \times
  \rme^{- \sum_k \rmi\hat\psi_k \mbox{\scriptsize \sc{Eq}}_{k-1} + \frac{1}{2} {\beta^{-1}} \sum_{kl} \rmi\hat\psi_k \Gamma_{kl} \rmi\hat\psi_l + \ln P_\mathrm{i}\left(\psi_1,\frac{\psi_1-\psi_0}{\epsilon}\right)}\; \nonumber
\end{eqnarray}
that in the continuum limit becomes
\begin{eqnarray}
{\cal N}  P[\psi]  &=&  |{\cal J}[\psi]| \, \rme^{\ln P_\mathrm{i}}  \int \uD{[\hat\psi]} \rme^{-\int \Ud{u} \rmi
\hat\psi(u) \mbox{\scriptsize \sc{Eq}}([\psi], u) +\frac{1}{2} \iint \Udd{u}{v}
    \rmi\hat\psi(u) {\beta^{-1}}\Gamma(u-v) \rmi\hat\psi(v) }
    \;, \nonumber
\end{eqnarray}
with the boundary conditions $\hat\psi(-T)=\hat\psi(T)=0$ and where
all the integrals over time run from $-T$ to $T$. In the following,
unless otherwise stated, we shall simply denote them by $\int$. The
infinite prefactor ${\cal N} \equiv \lim\limits_{N\to\infty} (2\pi/\epsilon)^N$
can be absorbed in the definition of the measure:
\begin{equation}
  {\cal D}[\psi,\hat \psi] = 
\lim\limits_{N\to\infty} \left(\frac{\epsilon}{2\pi}\right)^N\prod_{k=0}^{N+1} \ud{\psi_k}  \ud{\hat\psi_k} \;.
\end{equation}

\paragraph{Markov case.}\label{sec:MarkovianAction}
In the Markov limit, the Langevin equation is a first order differential equation, therefore only the 
first time $t_0$ should be reserved for integrating over the initial conditions. Moreover, one has to 
specify the discretization:
\begin{eqnarray}
\label{eq:MarkovLangevin}
\mbox{\sc{Eq}}_{k-1} &\equiv& 
\gamma_0  \frac{\psi_{k}-\psi_{k-1}}{\epsilon}
- F_{k}(\tilde\psi_{k}) = \xi_{k} 
\; , 
\end{eqnarray}
where $\tilde\psi_{k} \equiv a \psi_k + (1-a) \psi_{k-1}$ with $a \in [0,1]$. $a=0$ 
corresponds to the It\^o interpretation  whereas $a=1/2$ 
corresponds to the Stratonovich one (see the discussion in Sect.~\ref{sec:MarkovLim}). 
 Following 
the steps in \ref{app:MSRJD-Construct}, we upgrade eq.~(\ref{eq:MarkovLangevin}) to 
the following $a$-dependent action\footnote[1]{We omit the initial measure which is not relevant in this discussion.}:
\begin{eqnarray} \label{eq:ActionDiscretea}
\hspace{-1.3cm} S_N(a) =  \epsilon \sum_{k} \left( 
\beta^{-1} \gamma_0 (\rmi\hat\psi_k)^2 - \rmi\hat\psi_{k} \left[ \gamma_0 \frac{\psi_k - \psi_{k-1}}{\epsilon} -F_{k}(\tilde\psi_{k})  \right] 
 - \frac{a}{\gamma_0} F_{k}'(\tilde\psi_k) \right)
 \,.
\end{eqnarray}
The last term in the \textsc{rhs} comes from the Jacobian:
\begin{eqnarray*}
{\cal J}_N =
 \mbox{det}_{kl} \left(\frac{\partial \mbox{\sc{Eq}}_{k-1}}{\partial\psi_l} \right)
= \prod_k \left( \frac{\gamma_0}{\epsilon} - a F_k'(\tilde\psi_k) \right)
= \left(\frac{\gamma_0}{\epsilon}\right)^N  \rme^{- \epsilon \sum_k \frac{a}{\gamma_0} F_k'(\tilde\psi_k)}\;.
\end{eqnarray*}
In the It\^o discretization scheme ($a=0$) this Jacobian term disappears from the 
action. Although $S_N(a)$ seems to be $a$-dependent, we now prove that all  discretization 
schemes yield the same physics by showing that the difference $S_N(a)-S_N(0)$ is 
negligible. The Taylor expansion of $F_{k}(\tilde\psi_{k})$ around
$\psi_{k-1}$,  $F_{k}(\psi_{k-1}) + a \left(\psi_k - \psi_{k-1} \right) F'(\psi_{k-1}) + O(\epsilon)$
[since $\psi_k-\psi_{k-1}=O(\sqrt{\epsilon})$],
yields 
\begin{eqnarray}
S_N(a) - S_N(0) &=&  a \epsilon \sum_k F'(\psi_{k-1}) \left[  \rmi\hat\psi_k 
\left(\psi_k - \psi_{k-1} \right) - \frac{1}{\gamma_0} \right]
+ O(\epsilon^2)
\;.
\end{eqnarray}
Although the first term within the square brackets looks smaller than the second
one, they are actually both $O(1)$ since $i\hat \psi_k=O(1/\sqrt{\epsilon})$.
Thus, each term in the sum in the {\sc rhs} is $O(\epsilon)$. 
We now compute the average of $S_N(a) - S_N(0)$ with respect to 
$S_N(0)$ by neglecting in the latter the term $\epsilon\rmi\hat\psi_k F_k(\psi_{k-1})$ which is of 
order $\sqrt{\epsilon}$ whereas the others are of order $1$.
Since $\langle  \rmi\hat\psi_k \left(\psi_k - \psi_{k-1} \right) \rangle_{S_N(0)} = 1/\gamma_0$, it is easy to 
show that  $\langle 
S_N(a) - S_N(0) \rangle_{S_N(0)} =0$ and therefore  all the $S_N(a)$ actions are equivalent to 
the simpler It\^o one.

\subsection{\hspace{1.7cm}Jacobian} 
\label{app:Jacobian}

\subsubsection{\hspace{1.7cm}Discrete evaluation of the Jacobian.} 

In this section we take the continuum limit of the Jacobian defined
in eq.~(\ref{eq:jac_discr_def}).  In the additive noise case, we start from
\begin{eqnarray}
 {\cal J}_N &=& \mbox{det}  \left( \frac{\partial \, (\psi_\mathrm{i},\dot\psi_\mathrm{i}, \mbox{\small\sc{Eq}}_0, \ldots, \mbox{\small\sc{Eq}}_{N-1} )}{\partial \, (\psi_0,\psi_1,\ldots,\psi_{N+1})} \right) \nonumber \\
  &=& \mbox{det} \left( \begin{array}{cccccc}
			0 & 1 & 0\ldots & & &  \\
			-1/\epsilon & 1/\epsilon & 0\ldots & &  &\\
			\frac{\partial \mbox{\scriptsize \sc{Eq}}_0}{\partial\psi_0} & \frac{\partial \mbox{\scriptsize \sc{Eq}}_0}{\partial\psi_1}& \frac{\partial \mbox{\scriptsize \sc{Eq}}_0}{\partial\psi_2} & 0\ldots & \\
			\frac{\partial \mbox{\scriptsize \sc{Eq}}_1}{\partial\psi_0} &\frac{\partial \mbox{\scriptsize \sc{Eq}}_1}{\partial\psi_1} &\frac{\partial \mbox{\scriptsize \sc{Eq}}_1}{\partial\psi_2}& \frac{\partial \mbox{\scriptsize \sc{Eq}}_1}{\partial\psi_3}  & 0\ldots & \\
			\ldots & & & &  & 0 \\
 			\frac{\partial \mbox{\scriptsize \sc{Eq}}_{N-1}}{\partial\psi_{0}} & & \ldots & & &\frac{\partial \mbox{\scriptsize \sc{Eq}}_{N-1}}{\partial\psi_{N+1}}
                    \end{array}
\right) \nonumber \\
&=& \frac{1}{\epsilon}
\mbox{det} \left( \begin{array}{cccc}
			 \frac{\partial \mbox{\scriptsize \sc{Eq}}_0}{\partial\psi_2} & 0\ldots & \\
			\frac{\partial \mbox{\scriptsize \sc{Eq}}_1}{\partial\psi_2}& \frac{\partial \mbox{\scriptsize \sc{Eq}}_1}{\partial\psi_3}  & 0\ldots & \\
			\ldots &  &  & 0\\
 			\frac{\partial \mbox{\scriptsize \sc{Eq}}_{N-1}}{\partial\psi_{2}} & \ldots & &\frac{\partial \mbox{\scriptsize \sc{Eq}}_{N-1}}{\partial\psi_{N+1}}
                    \end{array}
\right)\;. \label{eq:detdisc}
\end{eqnarray}
Causality manifests itself in the lower triangular structure of the last matrix. 
One can evaluate the last determinant by plugging eq.~(\ref{eq:Langevin_discr}).
It yields
\begin{eqnarray}
 {\cal J}_N &=& \frac{1}{\epsilon} \prod_{k=1}^{N} \frac{\partial\mbox{\small \sc{Eq}}_{k-1}}{\partial\psi_{k+1}} = \frac{1}{\epsilon} \left( \frac{m}{\epsilon^2} \right)^N \;. \nonumber
\end{eqnarray}
The Jacobian ${\cal J} \equiv \lim\limits_{N\to\infty} {\cal J}_N$ is therefore a
field-independent positive constant that can be absorbed in a redefinition of the measure:
\begin{equation}\label{eq:defmes}
  {\cal D}[\psi,\hat \psi] \equiv \lim\limits_{N\to\infty}  \frac{1}{\epsilon} \left(\frac{m}{2\pi\epsilon}\right)^N\prod_{k=0}^{N+1} \ud{\psi_k}  \ud{\hat\psi_k} \;.
\end{equation}
 We show that this result also holds for 
multiplicative noise in \ref{app:MSRJDmult}.

\subsubsection{\hspace{1.7cm}Continuous evaluation of the Jacobian.}
\label{app:Jacobian1.2}

One might also wish to check this result in the continuous notations.
A very similar approach can be found in~\cite{Ohta}. In the continuous
notations, $\lim\limits_{N\to\infty}{\cal J}_N $ reads up to some constant factor
\begin{eqnarray*}
 {\cal J}[\psi]  = \mbox{det}_{uv} \left[
\frac{\delta\mbox{\small \sc{Eq}}([\psi],u)}{\delta \psi(v)} \right]\;.
\end{eqnarray*}
where $\mbox{det}\left[ ... \right]$ stands for the functional determinant. Defining $F'_{uv}$ as  $\delta F_u[\psi] / \delta \psi_v $, 
the Jacobian reads
\begin{eqnarray}
\hspace{-2.0cm}  {\cal J}[\psi]  = \mbox{det}_{uv} \left[ 
m\partial^2_u \delta_{u-v} + \int_w\gamma_{u-w} \, \partial_w \delta_{w-v} -   F_{uv}'[\psi]
 \right] \nonumber \\
\hspace{-1.0cm} = \mbox{det}_{uv}  \left[ m\partial_u^2 \delta_{u-v} + \int_w\gamma_{u-w} \, \partial_w \delta_{w-v}  \right]   \,  \mbox{det}_{uv}  \left[ \delta_{u-v} - \int_w G_{u-w} F'_{wv}[\psi]  \right]  \; \nonumber \\
\hspace{-1.0cm} = \mbox{det}_{uv} \left[ m\partial_u^2 \delta_{u-v} + \int_w\gamma_{u-w} \, \partial_w \delta_{w-v}  \right]   \,  \exp \mbox{Tr}_{uv}  \ln \left[ \delta_{u-v} - M_{uv}  \right] \; \nonumber \\
\hspace{-1.0cm} = \mbox{det}_{uv}  \left[ m\partial_u^2 \delta_{u-v} + \int_w\gamma_{u-w} \, \partial_w \delta_{w-v}  \right]   \,  \exp -\sum_{n=1}^{\infty}  \frac{1}{n} \int_u  \left\{ \underbrace{M {\circ} M {\circ} ... {\circ}M}_{n \mbox{ \scriptsize times}} \right\}_{uu} \label{eq:sum_tr2}
\end{eqnarray}
where we used the notations  $M_{uv} \equiv \left\{ G\circ F'\right\}_{uv} \equiv \int_w G_{u-w} F'_{wv}[\psi]$.
$G$ is the retarded Green function solution to
\begin{equation}
 m\partial^2_u G(u-v)  + \int \ud{w} \gamma(u-w) \partial_w  G(w-v)= \delta(u-v)\;. 
 \label{eq:GreenGen}
\end{equation}
Since both $G_{u-v}$ and $F'_{uv}$ are causal, it is easy to see that
the $n\geq2$ terms do not contribute to the sum in eq.~(\ref{eq:sum_tr2}).
If the force $F([\psi],t)$ does not have any local term (involving the value of $\psi$ or $\dot\psi$ at 
time $t$) the $n=1$ term is also zero. Otherwise the $n=1$ term can still be proven to be zero 
provided that  $G(t=0) = 0$. This will be true, as we shall show in the next paragraph,  unless the 
white noise limit is taken together with the Smoluchowski limit ($m=0$). Away from this Markov 
limit we establish
\begin{eqnarray}
 {\cal J}[\psi]  = \mbox{det}_{uv} \left[m\partial^2_u \delta(u-v) + \int_w \gamma_{u-w} \, \partial_w \delta_{w-v} \right]  \;, \nonumber
\end{eqnarray}
meaning that the Jacobian is a constant that does not depend on the field $\psi$.

We now give a proof that $G(t=0)=0$.
Taking the Fourier transform of eq.~(\ref{eq:GreenGen}), 
\begin{equation}
 G(t=0) = \int_{-\infty}^{\infty} \frac{\udm{\omega}}{2\pi} G(\omega)
  = -\int_{-\infty}^{\infty} \frac{\udm{\omega}}{2\pi} \frac{1}{m\omega^2 + 
  \rmi \omega \gamma(\omega)}
 \; . \label{eq:G0}
\end{equation}
$G(\omega)$ and $\gamma(\omega)$ are the Fourier transforms of the
retarded Green function and friction. They are both
analytic in the upper half plane ({\sc uhp}) thanks to their causality
structure.  The convergence of the integrals around
$|\omega|\to\infty$ in eq.~(\ref{eq:G0}) is ensured by either the
presence of inertia or the colored noise. For a white
noise [$\gamma(\omega) = \gamma_0$], it is clear that the mass term
renders the integrals in eq.~(\ref{eq:G0}) well defined. In the
$m=0$~limit the convergence is still guaranteed as long as the white
noise limit is not taken simultaneously. Indeed, because
$\gamma(\omega)$ is analytic in the {\sc {\sc uhp}}, it is hence either divergent
on the boundaries of the {\sc uhp} or constant everywhere
[$\gamma(\omega)=\gamma_0$]. In the first case, which corresponds to a
generic colored noise, this renders the integrals in eq.~(\ref{eq:G0})
well defined. In the second case, corresponding to a white noise
limit, they are ill-defined and require a more careful
treatment\footnote[1]{ In the white noise limit, $G(t)=
  \gamma_0^{-1}\left[1-\rme^{-\gamma_0 t/m} \right]\Theta(t)$ is a
  continuous function that vanishes at $t=0$. If we take $m\to0$ in
  the previous expression, we still have $G(0)=0$ and $G(t) =
  \Theta(t)/\gamma_0$ for $t\gg m/\gamma_0$. By choosing
  $\Theta(0)=0$, these two results can be collected in $G(t) =
  \Theta(t)/\gamma_0$ for all $t$.  The Jacobian is still a
  constant. This limiting procedure where inertia has been sent to
  zero after the white noise limit was taken, is the so-called It\^o
  convention. However if $m$ is set to $0$ from the beginning, in the
  so-called Stratonovich convention with $\Theta(0) = 1/2$, then $G(t)
  = \Theta(t)/\gamma_0$ for all $t$ and $G(0) = 1/(2\gamma_0)$. This
  can lead to a so-called Jacobian extra-term in the action. If
  $F([\psi],t)$ is a function of $\psi(t)$ only (ultra-local
  functional), it reads $-{1}/{(2\gamma_0)} \, \int_u
  F_u'(\psi_u)$. It is invariant under time-reversal of the field
  $\psi_u\mapsto\psi_{-u}$ as long as $F'$ is itself time-reversal
  invariant.}.  When the integrals in eq.~(\ref{eq:G0}) are well
defined on the boundaries, the absence of poles (or branch cuts) in the
{\sc uhp} of $G(\omega)$ gives, after a little deformation of the
integration contour in eq.~(\ref{eq:G0}) above the $\omega=0$ pole,
the result $G(t=0)=0$.

\subsubsection{\hspace{1.7cm}Representation in terms of a fermionic field integral.}
\label{app:Jacobian2}

The determinant can be represented as  a Gaussian integration
over Grassmannian conjugate fields $c$ and $c^*$.  This formulation is
a key ingredient to the supersymmetric representation of the MSRJD
path integral. Let us first recall the discretized expression of the
Jacobian obtained in eq.~(\ref{eq:detdisc}):
\begin{eqnarray*}
 {\cal J}_N =  \frac{1}{\epsilon} \mbox{det}_{kl} \left( \frac{\partial\mbox{\small\sc{Eq}}_{k-1}}{\partial\psi_{l+1}} \right) \;,
\end{eqnarray*}
where $k$ and $l$ run from $1$ to $N$. Introducing ghosts, it can be put in the form
\begin{eqnarray}
 {\cal J}_N &=&  \frac{1}{\epsilon} \frac{1}{\epsilon^N} \int \Udm{c_2}\udm{c_0^*} ... \udm{c_{N+1}}\udm{c_{N-1}^*} \, \rme^{ \epsilon^2 \sum_{k=0}^{N-1} \sum_{l=2}^{N+1} \, c_k^*\,  \frac{1}{\epsilon}\frac{\partial\mbox{\tiny\sc{Eq}}_{k}}{\partial\psi_{l}} \, c_l }  
 \nonumber\\
&=& \frac{1}{\epsilon} \frac{1}{\epsilon^{N}} \int \Udm{c_0}\udm{c_0^*} ... \udm{c_{N+1}}\ud{c_{N+1}^*}  c_{N+1}^*  c_N^* c_1 c_0   \,  \rme^{  \epsilon^2 \sum_{k=0}^{N+1} \sum_{l=0}^{N+1}  \, c_k^*\,  \frac{1}{\epsilon}\frac{\partial\mbox{\tiny\sc{Eq}}_{k}}{\partial\psi_{l}} \, c_l }\,, \nonumber
\end{eqnarray}
where in the last step, we allowed integration over $c_0$, $c_1$, $c_{N}^*$ and $c_{N+1}^*$ at the cost of introducing delta functions (remember that for a Grassmann number $c$, the delta function is achieved by $c$ itself).
In the continuum limit, absorbing the prefactor into a redefinition of the measure, 
\begin{equation}\label{eq:defmes2}
 \hspace{-3em}{\cal D}[\psi,\hat \psi] = \lim\limits_{N\to\infty} \frac{1}{(2\pi)^N} \frac{1 }{\epsilon} \prod_{k=0}^{N+1} \ud{\psi_k}  \ud{\hat\psi_k} \; \mbox{ and }\; \uD{[c,c^*]} = \lim\limits_{N\to\infty} \prod_{k=0}^{N+1} \ud{c_k}  \ud{c^*_k}\;,
\end{equation}
this yields
\begin{eqnarray*}
 {\cal J}[\psi]   =  \int \uD{[c,c^*]} \rme^{S^{\cal J}[c,c^*,\psi]}\;
\end{eqnarray*}
with
\begin{eqnarray*}
 S^{\cal J}[c,c^*,\psi] \equiv {\int_u \int_v c^*_u\, \frac{\delta\mbox{\small\sc{Eq}}_u[\psi]}{\delta \psi_v} \,  c_v} \;,
\end{eqnarray*}
and the extra boundary conditions:  $c(-T) = \dot c(-T) = c^*(T)  = \dot c^*(T) = 0$.
Plugging the Langevin equation~(\ref{eq:Langevin}), we have
\begin{eqnarray}
 \frac{\delta\mbox{\small\sc{Eq}}_u[\psi]}{\delta \psi_v} &=& m\partial^2_u \delta_{u-v} - \frac{\delta F_u[\psi]}{\delta \psi_v}  + \int_w \gamma_{w-v} \partial_w \delta_{w-v} \;. \nonumber
\end{eqnarray}
The kinetic term in $S^{\cal J}[c,c^*,\psi]$ can be re-written
\begin{eqnarray}
 \int_u\int_v  c^*_u \, \partial^2_u \delta_{u-v} \, c_v 
= \int_u c^*_u \, \partial^2_u c_u
 + \Theta_0 \left[ \dot c^* c - c^* \dot c \right]_{-T}^{T} 
 + \Theta_0 \delta_0 \left[ c^* c \right]_{-T}^{T} \;. \nonumber
\end{eqnarray}
The last two terms in the {\sc rhs} vanish by use of the boundary conditions 
($c_{-T}= \dot c_{-T} = c^*_{T} = \dot c^*_{T}=0$).
The retarded friction can be re-written
\begin{eqnarray}
 \int_u \int_v c^*_u \, \partial_u \gamma_{u-v} \, c_v
 - \Theta_0 \int_u c^*_u \left[ \gamma_{u+T} \, c_{-T}  - \gamma_{u-T} \, c_{T} \right] \;, \nonumber
\end{eqnarray}
where the second term vanishes identically for two reasons: the boundary condition ($c_{-T}=0$) kills the first part and the causality of the friction kernel ($\gamma_u = 0 \, \forall\, u  < 0$) 
suppresses the second one. If there is a Dirac contribution to $\gamma$ centered at $u=0$ like in 
the white noise case, the other boundary condition ($c^*_{-T}=0$)  cancels the second 
part. Finally, we have
\begin{eqnarray}
S^{\cal J}[c,c^*,\psi]  = \int_u c^*_u \, \partial^2_u c_u + \int_u \int_v c^*_u \left[\partial_u \gamma_{u-v} -  \frac{\delta F_u[\psi]}{\delta \psi_v} \right] c_v\;.
\end{eqnarray}

\section{\hspace{1.8cm} Discrete MSRJD for multiplicative noise}\label{app:MSRJDmult}
The discretized Langevin equation reads:
\begin{eqnarray*}
\mbox{\textsc{Eq}}_{k-1} &\equiv& 
m \frac{\psi_{k+1} - 2 \psi_{k}  + \psi_{k-1} }{\epsilon^2}  - {F_{k}}(\tilde\psi_{k}, \tilde\psi_{k-1}, ...) \nonumber \\
& & \qquad  + M'(\tilde\psi_{k}) \, \epsilon\sum_{l=1}^{k}  \gamma_{kl}  M'(\tilde\psi_l)  \frac{\psi_{l}-\psi_{l-1}}{\epsilon} 
= M'(\tilde\psi_{k}) \xi_{k} \;. 
\end{eqnarray*}
with $\tilde\psi_{k} \equiv a \psi_k + (1-a) \psi_{k-1}$ and 
$k=1, ..., N$. In the Markov limit ($m=0$ and $\gamma_{kl} = \epsilon^{-1} \gamma_0 \delta_{kl}$) 
the results depend on $a$ (see the discussion in Sect.~\ref{sec:MarkovLim}). In the 
additive noise case, the choices $a=0$ and $a=1/2$ correspond to the It\^o and 
Stratonovich conventions, respectively.  However, we decide to stay out of the Markov limit: the 
results are then independent of $a$ and we choose to work with $a = 1$. The probability for a field 
history is 
\begin{eqnarray} 
 P(\psi_0, \psi_1, ..., \psi_{N+1}) = |{\cal J}_N| P_\mathrm{i}(\psi_1, \frac{\psi_1 - \psi_0}{\epsilon}) P_\mathrm{n}(\widetilde{\mbox{\small \sc{Eq}}}_0, ..., \widetilde{\mbox{\small \sc{Eq}}}_{N-1}) \;,
\end{eqnarray}
where we introduced the shorthand notation $\widetilde{\mbox{\small\sc{Eq}}}_k \equiv \mbox{\small\sc{Eq}}_k / M'(\psi_{k+1})$. 
The Jacobian is
\begin{eqnarray}
 \hspace{-1.5em} {\cal J}_N \equiv \mbox{det}  \left( \frac{\partial\,(\psi_\mathrm{i},\dot\psi_\mathrm{i}, \xi_1, \ldots, \xi_N )}{\partial\, (\psi_0,\psi_1,\ldots,\psi_{N+1})} \right)
= \mbox{det} \left( \frac{\partial \, (\psi_\mathrm{i},\dot\psi_\mathrm{i}, \widetilde{\mbox{\small\sc{Eq}}}_0, \ldots, \widetilde{\mbox{\small\sc{Eq}}}_{N-1} )}{\partial \, (\psi_0,\psi_1,\ldots,\psi_{N+1})} \right). \label{eq:jacdiscmult}
\end{eqnarray}
$P_\mathrm{n}$ is still given by expression~(\ref{eq:noisediscreteHubb}) and 
$P_\mathrm{n}(\widetilde{\mbox{\small \sc{Eq}}}_0, ..., \widetilde{\mbox{\small \sc{Eq}}}_{N-1})$ reads,
 after the substitution $\hat\psi_k  \mapsto \hat\psi_k  M'(\psi_k) $,
\begin{eqnarray*}
\hspace{-1.5cm} {\cal N}_N^{-1} \int \Udm{\hat\psi_0}...\ud{\hat\psi_{N+1}} \delta(\hat\psi_0)\delta(\hat\psi_{N+1})\, |\hat{\cal J}_N| \, \rme^{ - \epsilon\sum_k \rmi\hat\psi_k \mbox{\scriptsize \sc{Eq}}_{k-1} + \frac{1}{2} {\beta^{-1}} \epsilon^2\sum_{kl}  \rmi\hat\psi_k M'(\psi_k) \Gamma_{kl} M'(\psi_l) \rmi\hat\psi_l}  \;, \nonumber 
\end{eqnarray*}
where $\hat{\cal J}_N\equiv \mbox{det}_{kl} \left({\delta_{k\,l}} \, {M'(\psi_{k})} \right) $ is the Jacobian of the previous substitution.
The probability for a given history is therefore
\begin{eqnarray*}
 P(\psi_0, \psi_1, ..., \psi_{N+1})  = {\cal N}_N^{-1}   \int \Udm{\hat\psi_0}...\ud{\hat\psi_{N+1}}  \left|{\cal J}_N \hat{\cal J}_N\right| \\
\qquad\qquad\times
 \rme^{- \sum_k \rmi\hat\psi_k \mbox{\scriptsize \sc{Eq}}_{k-1} + \frac{1}{2} {\beta^{-1}} \sum_{kl} \rmi\hat\psi_k M'(\psi_k) \Gamma_{kl} M'(\psi_l) \rmi\hat\psi_l + \ln P_\mathrm{i}\left(\psi_1,\frac{\psi_1-\psi_0}{\epsilon}\right)}\;. \nonumber
\end{eqnarray*}
The Jacobian  ${\cal J}_N$ defined in eq.~({\ref{eq:jacdiscmult}}) reads
\begin{eqnarray}
 {\cal J}_N &=&  \frac{1}{\epsilon} \, \mbox{det}_{kl} \left(\frac{1}{M'(\psi_{k})}\frac{\partial\mbox{\small\sc{Eq}}_{k-1}}{\partial\psi_{l+1}} 
- \frac{M''(\psi_{k})}{M'(\psi_{k})^2} \, \mbox{\small\sc{Eq}}_{k-1} \,\delta_{k \, l+1 } \right) \nonumber \\
&=&  \frac{1}{\epsilon} \, \hat{\cal J}_N^{-1} \,
\mbox{det}_{kl} \left(\frac{\partial\mbox{\small\sc{Eq}}_{k-1}}{\partial\psi_{l+1}} 
- \frac{M''(\psi_{k})}{M'(\psi_{k})} \, \mbox{\small\sc{Eq}}_{k-1} \,\delta_{k \, l+1 } \right) \label{eq:jacdismult}
\end{eqnarray}
where $k$ and $l$ run from $1$ to $N$. 
Causality is responsible for the triangular structure of the matrix involved in the last expression. 
The second term within the square brackets yields matrix elements below the main 
diagonal and these do not contribute to the Jacobian.
Therefore, we find 
\begin{eqnarray}
 {\cal J}_N \hat{\cal J}_N =  \frac{1}{\epsilon} \prod_{k=1}^{N} \frac{\partial\mbox{\small \sc{Eq}}_{k-1}}{\partial\psi_{k+1}} = \frac{1}{\epsilon} \left( \frac{m}{\epsilon^2} \right)^N \;. \nonumber
\end{eqnarray}
that is the same field-independent positive constant as in the additive noise case that can be dropped in the measure, see eq.~(\ref{eq:defmes}).

A fermionic functional representation of the Jacobian can be obtained
by introducing ghosts,  expression~(\ref{eq:jacdismult}) can be put in the form
\begin{eqnarray}
 {\cal J}_N \hat{\cal J}_N &=& \frac{1}{\epsilon} \frac{1}{\epsilon^N} \int \udm{c_0}\udm{c_0^*} ... \udm{c_{N+1}}\ud{c_{N+1}^*}  c_{N+1}^*c_N^* c_1 c_0 \  \rme^{S^{\cal J}_N} \;,\nonumber 
\end{eqnarray}
with 
\begin{eqnarray*}
S^{\cal J}_N &\equiv& \epsilon^2 \sum_{k=0}^{N+1} \sum_{l=0}^{N+1}  \, c_k^*\,  \frac{1}{\epsilon}\frac{\partial\mbox{\small \sc{Eq}}_{k}} {\partial\psi_{l}} \, c_l
-\epsilon \sum_{k=0}^{N+1} c_k^* \, \frac{M''(\psi_{k+1})}{M'(\psi_{k+1})} \, \mbox{\small\sc{Eq}}_{k} \, c_{k+1} \;.
\end{eqnarray*}
In the continuum limit it becomes
\begin{eqnarray*}
  S^{\cal J} &\equiv& \lim\limits_{N \to \infty} S^{\cal J}_N = {\int_u \int_v  c^*_u\, \frac{\delta\mbox{\small\sc{Eq}}_u[\psi]}{\delta \psi_v} \,  c_v} - \int_u c^*_u \, \frac{M''(\psi_u)}{M'(\psi_u)} \mbox{\small\sc{Eq}}_u[\psi] \, c_u \;,
\end{eqnarray*}
with the boundary conditions $c(-T) = \dot c(-T) = 0$ and $c^*(T) = \dot c^*(T) = 0$ and the measure of the corresponding path integral is given in (\ref{eq:defmes2}).

\end{document}